%% file: article_draft_final.tex
\documentclass[epj,iicol,numbook]{svjour}
\usepackage{authblk}
\providecommand{\newblock}{}
\usepackage{graphicx}
\usepackage{amsmath,amssymb}
\usepackage[square,numbers,compress]{natbib}
\usepackage{enumerate}
\usepackage[shortlabels]{enumitem}
\usepackage{url}
\usepackage[thinc]{esdiff}
\usepackage[dvipsnames]{xcolor}
\usepackage{setspace}
\usepackage{dutchcal}
\usepackage{mathtools}
\usepackage{tabularx}
\usepackage{scalerel}
\usepackage{braket}
\usepackage{tikz}
\usepackage{tikz-cd}
\usetikzlibrary{svg.path}
\usepackage{comment}
\usepackage{moreenum}
\usepackage[utf8]{inputenc}
\usepackage{bm}
\usepackage{slashed}
\usepackage{xparse}
\usepackage{dsfont}
\usepackage[mathscr]{euscript}
\usepackage{float}
\usepackage{cuted}
\usepackage[colorlinks,citecolor=blue,urlcolor=blue,linkcolor=blue]{hyperref}

\usepackage{microtype}

\newcommand{\p}{\partial}

\renewcommand{\d}{\text{d}}

\def\tl{\textrm{T}}
\def\td{{\cal T}}
\newcommand{\BK}{Brown-Kucha\v{r} }

\DeclarePairedDelimiter{\abs}{\lvert}{\rvert}

\labelformat{equation}{Eq.~(#1)} 
\labelformat{figure}{Fig.~#1} 
\labelformat{subfigure}{Fig.~\thefigure#1} 
\labelformat{table}{Tab.~#1} 
\labelformat{appendix}{Appendix #1}

\definecolor{orcidlogocol}{HTML}{A6CE39}
\tikzset{
  orcidlogo/.pic={
    \fill[orcidlogocol] svg{M256,128c0,70.7-57.3,128-128,128C57.3,256,0,198.7,0,128C0,57.3,57.3,0,128,0C198.7,0,256,57.3,256,128z};
    \fill[white] svg{M86.3,186.2H70.9V79.1h15.4v48.4V186.2z}
                 svg{M108.9,79.1h41.6c39.6,0,57,28.3,57,53.6c0,27.5-21.5,53.6-56.8,53.6h-41.8V79.1z M124.3,172.4h24.5c34.9,0,42.9-26.5,42.9-39.7c0-21.5-13.7-39.7-43.7-39.7h-23.7V172.4z}
                 svg{M88.7,56.8c0,5.5-4.5,10.1-10.1,10.1c-5.6,0-10.1-4.6-10.1-10.1c0-5.6,4.5-10.1,10.1-10.1C84.2,46.7,88.7,51.3,88.7,56.8z};
  }
}

\newcommand\orcidicon[1]{\href{https://orcid.org/#1}{\mbox{\scalerel*{
\begin{tikzpicture}[yscale=-1,transform shape]
\pic{orcidlogo};
\end{tikzpicture}
}{|}}}}

\DeclareUnicodeCharacter{2212}{-}

\begin{document}

\title{Unitary quantum matter-bounce in a universe with a positive cosmological constant\thanks{Dedicated to the memory of Prof.~Jayant V.~Narlikar}}

\titlerunning{Unitary quantum matter-bounce in a universe with a positive $\Lambda$}

\author{Harkirat Singh Sahota\thanks{harkirat221@gmail.com}\and Dipayan Mukherjee\thanks{dipayanmkh@gmail.com} \and S.~Shankaranarayanan\thanks{shanki@iitb.ac.in}}
\institute{Department of Physics, Indian Institute of Technology Delhi, Hauz Khas, New Delhi, 110016, India. \and
Raman Research Institute, C.~V.~Raman Avenue, Sadashivanagar, Bengaluru 560080, India.\and
Department of Physics, Indian Institute of Technology Bombay, Mumbai 400076, India.
}
\authorrunning{Sahota\and Mukherjee\and Shankaranarayanan}

\date{July 2026}

\abstract{We analyze the Wheeler-DeWitt quantization of a spatially flat Friedmann-Lemaître-Robertson-Walker universe containing pressureless dust and a positive cosmological constant ($\Lambda > 0$). Following relational time framework, we establish a direct mathematical correspondence between the cosmological Hamiltonian and the radial Schr\"odinger equation for the scattering states of the non-relativistic hydrogen atom. This exact solvability allows us to rigorously construct the physical Hilbert space and ensure the self-adjointness of the Hamiltonian. As a concrete result, we show that the wave packets unitarily evolve depicting a non-singular quantum bounce, systematically replacing the classical Big Bang singularity. Finally, we discuss the physical relevance of this exact solution within the matter-bounce scenario. We demonstrate that this framework provides a robust quantum origin for a bounce during a dust-dominated contracting phase --- a necessary prerequisite for generating a scale-invariant spectrum of primordial perturbations --- derived from the unitary dynamics of the quantized background.}

\maketitle


\section{Introduction}\label{Sec1}

Classical general relativity, a remarkably successful theory, famously predicts the existence of spacetime singularities where curvature diverges and the theory itself breaks down, like singularities within black holes and the initial Big Bang in cosmology~\cite{Penrose1965,Hawking:1971vc}. At such extreme densities and curvatures, quantum nature of gravity is expected to play a crucial role~\cite{Padmanabhan:1982bd,Narlikar:1986kr,Kiefer2012-dq}. The question of quantization of gravity can be succinctly cast within simplified, symmetry-reduced minisuperspace models, which serve as tractable theoretical frameworks for exploring the quantum gravitational effects. Attempts to explore quantum effects in the early universe have been made across various approaches, including Wheeler-DeWitt (WDW) quantization \cite{Thebault:2022dmv}, Loop Quantum Cosmology \cite{bojowald_loop_2008,Agullo2023}, and aspects of string cosmology, such as the ekpyrotic universe \cite{Khoury:2001wf,Buchbinder:2007at}.

The quantization of Friedman universes with a perfect fluid in WDW framework has been extensively investigated~\cite{Alvarenga:1998wx,Alvarenga:2001nm,kiefer_classical_2006,kreienbuehl_singularity_2009,BarberoG:2010oga,Husain:2011tm,Husain:2011tk,Bergeron:2013ika,Bergeron:2014kea,liu_singularity_2014,Malkiewicz:2014fja,bergeron_singularity_2015,Malkiewicz:2017iuj,kiefer_singularity_2019,kiefer_singularity_2019-1,Malkiewicz:2019azw,Husain:2019nym,Sahota_Infrared,Malkiewicz:2022szx,Mukherjee:2023qan,Sahota2023,Sahota:2023kox,Gielen_PRL,Piazza:2025fbt,Vitenti:2026hgs}. However, the quantization beyond single fluid becomes analytically intractable except for the case of a massless scalar field in a perfect fluid dominated universe \cite{Gryb:2018whn,Gielen:2020abd,Gielen:2021igw,Gielen:2022tzi,Alexandre:2022npo} that retain the mathematical structure of a single fluid quantum model. In this work, we investigate the WDW quantization of a Friedman universe with cosmological constant and dust. Earlier investigations of this system are reported in \cite{Amemiya_2009,maeda_unitary_2015,Ali:2018vmt}, where the quantum system is numerically evolved with the dust variable as the clock choice. The case of unimodular dynamics of fluid-$\Lambda$ system is undertaken in \cite{Gielen:2022dhg,Alexandre:2022ijm}, where the WDW equation is solved in the \emph{connection-representation}, transforming it into a first-order differential equation. This method allows for analytical investigation, but the underlying mathematical framework and the interpretation of states, spectrum and observables are distinct from a direct volume-representation that we consider in this work.

However, the principal challenge in quantizing general relativity stems from its nature as a gauge theory~\cite{Dirac:1958sc}. In the canonical formulation, spacetime diffeomorphism invariance manifests as gauge transformations in phase space, generated by constraint functions~\cite{Bergmann:1972ud}. The Hamiltonian constraint, in particular, enforces time-reparameterization invariance. Consequently, canonical quantization of such generally covariant systems typically yields the \emph{timeless} WDW equation \cite{wheeler_superspace_nodate, dewitt_quantum_1967,Narlikar:1983lum}. This absence of an external time parameter, and thus an extrinsic notion of dynamics in the quantum theory, is a characteristic feature of generally covariant theories and constitutes the well-known \emph{problem of time} in quantum gravity \cite{Isham:1992ms, kuchar_time_2011, anderson_2012}.

To reintroduce a notion of dynamics in this timeless framework, one often adopts a \emph{relational approach}: the evolution of certain dynamical variables is described relative to another variable designated as a reference clock~\cite{Tambornino:2011vg, Hoehn:2019fsy, Chataignier:2021ncn}. A critical question then arises in the quantum picture: at what stage should the clock (or gauge) fixing occur? Specifically, whether one first quantizes the full constrained system and then extracts the unitary dynamics, or first fixes the gauge classically and then quantizes the reduced phase space~\cite{Giulini_1999, Chataignier:2019kof, Barvinsky:2013aya}. Furthermore, the choice of gauge, often equivalent to selecting a particular dynamical variable as the clock, is not unique ~\cite{Barvinsky:2013aya}, leading to the \emph{clock ambiguity} problem ~\cite{Hohn:2018toe}. Different clock choices can yield unitarily inequivalent quantum theories for the same physical system~\cite{Gotay_Singularity}, a situation sometimes interpreted as a potential breakdown of general covariance at the quantum level~\cite{Malkiewicz:2019azw,Gielen:2020abd}.

Keeping these issues in mind, this work focuses on the WDW quantization \cite{wheeler_superspace_nodate, dewitt_quantum_1967,Narlikar:1983lum} of a spatially flat Friedmann-Lema\^itre-Robertson-Walker (FLRW) cosmological model. The model incorporates pressureless dust, described using the \BK formalism~\cite{brown_dust_1995}, and a cosmological constant, introduced dynamically via the unimodular approach \cite{Unruh:1988in, Henneaux:1989zc}. We consider the unimodular variable ($\tl$) associated with the cosmological constant as the clock \cite{Unruh:1988in, Henneaux:1989zc, Kuchar:1991xd, Smolin:2009ti},
and show the quantization problem of the FLRW universe with dust and a cosmological constant has a \emph{remarkable resemblance to that of a hydrogen atom}, the quantum mechanics of which is well-explored. The primary novel contribution of this work is the exact, analytical solutions of WDW equation for a dust-filled universe with $\Lambda > 0$, leading to unitarily evolving bouncing universe.
This work primarily focuses on the implications for a positive cosmological constant ($\Lambda > 0$), while the detailed analysis of the negative cosmological constant sector ($\Lambda < 0$) is presented in \cite{QC-Letter}.

{The rest of this paper is organized as follows. In Sec.~\ref{Sec2} we discuss the canonical formulation of FLRW universe with \BK dust and unimodular treatment of cosmological constant. The quantization of this model is presented in Sec.~\ref{Sec3}, where we solve the WDW equation for this system. The unitarily evolving wave packets are constructed for this system in Sec.~\ref{Sec4} and we investigate the evolution of the quantum system through the probability density of the wave packet, expectation values, and quantum fluctuations of the scale factor. We highlight the significance of the present analysis in light of matter bounce scenario in Sec.~\ref{Sec5}. Finally, we summarize our findings in Sec.~\ref{Sec6}.}

Throughout this paper Greek indices represent spacetime components, and the metric signature is taken to be $(−, +, +, +)$. We set $\hbar = c = 1$ and and $\kappa^2 = 8 \pi G=1$. For further discussion on the choice of units and scales involved, see Appendix~\ref{Appendix_units}. An \emph{overdot} denotes derivative with respect to unimodular time (\textrm{t}) and a \emph{prime} stands for a derivative with respect to cosmic time ($\tau$).

\section{FLRW universe with dust and cosmological constant}\label{Sec2}

We begin with a brief review of the classical dynamics of a
spatially flat FLRW universe with a cosmological constant and
\BK dust. The Einstein-Hilbert action with \BK dust is~\cite{brown_dust_1995}
\begin{align}
\label{eq:action01}
  S = \frac{1}{2} \int \d^4 x \sqrt{-g} \left( \mathcal{R} - 2 \Lambda \right)- \frac{1}{2} \int \d^4 x \sqrt{-g} ~ \rho \left( g^{\alpha \beta}u_\alpha u_\beta + 1 \right),
\end{align}
where $\mathcal{R}$ is the Ricci scalar, $\Lambda$ is the
cosmological constant, $\rho$ and $u^\alpha$ are the energy density
and the four-velocity associated with \BK dust. 
The four-velocity of the \BK dust is parameterized as $u_\alpha=-\partial_\alpha \td+{\cal W}_a\partial_\alpha {\cal S}^a$ by \emph{seven non-canonical fields} $\td$, ${\cal W}_a$, and ${\cal S}^a$ with $a=1,\;2,\;3$ \cite{brown_dust_1995}.  
The fields $(\td(x),\;{\cal S}^a(x))$ provide the notion of dynamical reference frame whereas the field $\mathcal{W}_a(x)$ is a non-dynamical field that plays the role of Lagrange multiplier \cite{brown_dust_1995}. Owing to the homogeneity of the spatially flat FLRW spacetime,
\begin{align}
  \d s^2 = - N^2(t) \d t^2 + a^2(t) \delta_{ij} \d x^i \d x^j,
\end{align}
the Ricci scalar and the four-velocity of the \BK dust
take the following forms
\begin{align}
  \mathcal{R} &= \frac{6}{N^2} \left( \frac{1}{a^2}\left(\diff{a}{t}\right)^2 + \frac{1}{a}\frac{\d^2a}{\d t^2} -\frac{1}{Na} \diff{a}{t}\diff{N}{t} \right),\\
  u_\alpha & = - \delta_{\alpha}^t\diff{\td}{t}. 
\end{align}
In the cosmological minisuperspace approach, all non-canonical fields that describe the \BK dust vanish except $\td$ and, hence, can serve as the reference clock. At the classical level, the role of reference clock is redundant, however, as mentioned in the introduction,  quantum dynamics requires a well defined notion of clock \cite{Isham:1992ms,kuchar_time_2011,anderson_2012}. 

Interestingly, the notion of reference clock can be obtained from the cosmological constant under the framework of unimodular gravity~\cite{Kuchar:1991xd}. 
While $\Lambda$ is a constant in GR, it is a variable with additional constraint in the unimodular gravity framework~\cite{Henneaux:1989zc}. This is incorporated in the action \eqref{eq:action01} by introducing the dynamical fields ${\tl}^\alpha(x)$:
\begin{align}
  S_\Lambda &= \int \d^4x \Lambda\p_\alpha \tl^\alpha \, .
  \label{UniC}
\end{align}
Together with action \ref{eq:action01}, the above action 
acts as Lagrange multipliers that enforce the equation of motion for cosmological constant $\partial_\alpha\Lambda=0$ such that $\partial_\alpha\tl^\alpha=\sqrt{-g}$. The parametrized action leads to the unimodular formulation of general relativity, where the theory
is invariant under the transformations that leave the metric
determinant $\sqrt{-g}$ unchanged. For further discussion on different approaches to unimodular gravity, see~\cite{Smolin:2009ti}. Again, in the cosmological minisuperspace, only one unimodular variable survives which we will write as $\tl$. For the spatially flat-FLRW space-time, the action \ref{eq:action01} and \ref{UniC} (up to boundary terms) reduces to:
\begin{align}
  S =&\int \d t
      \biggr[ - \frac{3 a }{N}\left(\diff{a}{t}\right)^2
      -  N a^3 \Lambda+ \Lambda \diff{\tl}{t} \nonumber\\
       &\qquad
       -\frac{1}{2}  a^3 N\rho
      \left( - \frac{1}{N^2}\left(\diff{\td}{t}\right)^2 + 1 \right)
      \biggr] \, ,\label{action}
\end{align}
where the 3-D spatial volume is set to unity. We are interested in the quantization of this system, for which the Hamiltonian formulation of this system is the starting point. 

\subsection{Canonical formulation}\label{SubS2a}
In the canonical picture, the phase-space variables of the system
are $\{a , P_a, \tl, P_\tl, \td, P_\td,$ $ N, P_N\}$, where the
conjugate momenta $P_i$ are given by
\begin{align}
  P_a &= - \frac{6 a }{N}\diff{a}{t},
        \qquad P_\tl = \Lambda
        \label{momentum1}\\
  P_\td &= \frac{a^3 \rho}{N} ~\diff{\td}{t},
          \qquad P_N = 0.
          \label{momentum2}
\end{align}
We work with the volume variable, which is achieved by the canonical transformation $(a, P_a) \to (v, P_v)$
\begin{align}
  v &= a^3,\qquad
  P_v = \frac{P_a}{3 a^2}.
\end{align}
The total Hamiltonian with respect to the phase-space variables
$\{v , P_v, \tl, P_\tl, \td, P_\td, N, P_N\}$ then becomes
\begin{align}
  H = N \left( - \frac{3}{4} v P_v^2 + v P_\tl + P_\td \right).\label{HamCons}
\end{align}
The vanishing of the momentum conjugate to the lapse function is the primary constraint of the theory, and consistency of the primary constraint $\dot P_N=0$ leads to the secondary constraint of the theory $H\approx0$, referred to as the Hamiltonian constraint.

The Hubble parameter ($\mathcal{H}$) with general lapse function $N(t)$ takes the form
\begin{align}
  \label{Hubble_gen}
  \mathcal{H} \equiv
  \frac{1}{a}\frac{\d a}{N\d t}
  = \frac{1}{3  v}\frac{\d v}{N\d t},
\end{align}
It characterizes the evolution of the universe: the sign of the Hubble parameter, $\mathcal{H}$, distinguishes between an expanding ($\mathcal{H} > 0$) or contracting ($\mathcal{H} < 0$) universe. 
Depending on the expanding or contracting branch, the dynamics of this system start with or ends to a cosmological singularity, where the classical description fails and a quantum description is expected to dictate the near singularity behavior. 

\section{Quantization of dust dominated universe with cosmological constant}\label{Sec3}
The starting point for quantization of the universe is to promote phase space variables to operators in the Hamiltonian constraint in \ref{HamCons}. This leads to the WDW equation, which for the present system takes the form of a Schr\"{o}dinger-like equation
\begin{align}
\label{eq:WdW}
    \left[\frac{3v}{4}\frac{\partial^2}{\partial v^2}-iv\frac{\partial}{\partial \tl}-i\frac{\partial}{\partial \td}\right]\Psi(v, \td, \tl) = 0.
\end{align}
Note that here we have considered the trivial operator ordering of the kinetic term.
Recognizing the momentum conjugate to $\tl$ to be the cosmological constant $\Lambda$ and momentum conjugate to dust variable to be the energy density of dust $\rho_0$, we choose the following ansatz $\psi(v,\tl)=e^{i(\Lambda \tl+\rho_0\td)}\psi_{\Lambda,\rho_0}(v)$. With this ansatz, the system is in a representation where the momenta $P_{\tl}$ (identified with the cosmological constant $\Lambda$) and $P_{\td}$ (identified with the dust energy density $\rho_0$) are sharp. Substituting this in the above equation, we get: 
\begin{align}
   \frac{\d^2}{\d v^2}\psi_{\Lambda,\rho_0}(v) +\frac{4}{3v}(v\Lambda+\rho_0)\psi_{\Lambda,\rho_0}(v)=0.\label{WDW}
\end{align}
\emph{The timeless nature of the WDW equation} ($\hat{H}\Psi=0$) reflects the standard problem of time in quantum gravity~\cite{Isham:1992ms,kuchar_time_2011,anderson_2012}. To establish a meaningful notion of unitary evolution and construct the physical Hilbert space, one must identify an internal clock variable from within the system's dynamical variables. While both \BK and unimodular variable are viable choices, the unimodular clock emerges as a natural choice, as the corresponding Hamiltonian admits self-adjoint extensions. The issues with the \BK clock choice are detailed in Appendix~\ref{Appendix_unimodular}, and the self-adjointness of the unimodular Hamiltonian is established in Appendix~\ref{Appendix_self-adjoint}.

In this work, we focus exclusively on the positive cosmological constant sector ($\Lambda > 0$). This sector physically models a universe undergoing an early-time dust-dominated phase that transitions into a late-time de Sitter expansion. Analyzing the explicit solutions for $\psi_{\Lambda,\rho_{0}}(v)$ in this regime allows us to establish a direct mathematical correspondence with the continuous spectrum (scattering states) of the non-relativistic hydrogen atom. The properties of these exact solutions naturally inform the construction of the physical Hilbert space and the appropriate inner product for the cosmological wave functions. The general solutions to the WDW \ref{WDW} for a positive cosmological constant ($\Lambda > 0$) are given by:
\begin{align}
\begin{split}
    \psi^1_{\Lambda,\rho_0}(v)=v\;e^{-ikv}\mathcal{F} \left(1+i\frac{2\rho_0}{3k},2,2ikv\right),\\
    \psi^2_{\Lambda,\rho_0}(v)=v\;e^{-ikv}\;\mathcal{U} \left(1+i\frac{2\rho_0}{3k},2,2ikv\right),
\end{split}\label{PosSS}
\end{align}
where $k \equiv 2 \sqrt{\Lambda/3}$, $\mathcal{F}$ is the confluent hypergeometric function of the first kind (Kummer function), and $\mathcal{U}$ is the confluent hypergeometric function of the second kind (Tricomi function)~\cite{Abramowitz1965-ug}. 

\subsection{\texorpdfstring{$\Lambda$}{L}-dust analogy to hydrogen atom}\label{SubS3a}

This quantum system exhibits a striking mathematical parallel to the time-independent radial Schr\"odinger equation for the s-wave ($l=0$) states of a hydrogen atom. To see this clearly, recall that the radial Schr\"odinger equation for $l=0$, is given by~\cite{Zettili2009-wk}
\begin{equation}
\frac{\d^2u(r)}{\d r^2} + \left(\frac{2m_e E}{\hbar^2} + \frac{2m_e e^2}{4\pi\epsilon_0\hbar^2 r}\right)u(r) = 0
\end{equation}
A direct comparison shows the following analogies: First, the variable $v$ (related to the volume of the universe) plays the role of the radial coordinate $r$. Second, the term ${4}\Lambda/3$ is analogous to ${2m_e E}/{\hbar^2}$, establishing the cosmological constant $\Lambda$ as the counterpart to the energy $E$ of the hydrogen atom. As $\Lambda$ plays the role of energy, its conjugate momentum emerge naturally as a clock variable. Third, the term ${4\rho_0}/{3v}$ corresponds to the Coulomb interaction term ${2m_e e^2}/{(4\pi\epsilon_0\hbar^2 r)}$. Thus, the dust energy density parameter $\rho_0$ is proportional to the ``effective charge" determining the strength of the $1/v$ potential, specifically $\frac{4\rho_0}{3} \leftrightarrow \frac{2m_e e^2}{4\pi\epsilon_0\hbar^2}$. Lastly, the wave function $\psi_{\Lambda,\rho_0}(v)$ corresponds to $u(r)$.

The positive $\Lambda$ sector of this quantum system is analogous to the continuous spectrum (scattering states) of the hydrogen atom where $E > 0$. The parameter $k = 2\sqrt{\Lambda/3}$ is real and acts like a wave number. The solutions $\psi^1_{\Lambda,\rho_0}(v)$ and $\psi^2_{\Lambda,\rho_0}(v)$ in~\ref{PosSS}, involving confluent hypergeometric functions $\mathcal{F}$ and $\mathcal{U}$ with complex arguments, are analogous to Coulomb wave functions~\cite{Landau1981-ej}. These represent oscillatory states. For the continuous spectrum, the generalized eigenstates for $\psi^1_{\Lambda,\rho_0}(v)$ are delta-function normalized (see Appendix~\ref{Appendix_Norm}),
\begin{align}
    \psi_{k}^1(v)=\sqrt{\frac{4\rho_0e^{\frac{2\pi\rho_0}{3k}} }{3k\sinh\left(\frac{2\pi\rho_0}{3k}\right)}}v\;e^{-ik v}\mathcal{F} \left(1+i\frac{2\rho_0}{3k},2,2ik v\right),\label{SAPosSS}
\end{align}
where $k \equiv 2\sqrt{\Lambda/3}$ is the wave number associated with the $v$-coordinate for $\Lambda>0$. These stationary states form a complete, orthonormal basis (in the generalized sense~\cite{Madrid_2005}):
\begin{align}
    \braket{\psi_k|\psi_{k'}}=\delta(k-k').
\end{align}
These states will be used for the construction of unitarily evolving wave packets, enabling the study of the quantum dynamics of a dust-dominated universe with a cosmological constant. Notice, we are treating $P_\td$ as a c-number in this framework and we are not taking superposition over its eigenvalue $\rho_0$.

In this work, we focus on case of the positive cosmological constant sector of the quantum model, corresponding to a universe with an early time dust-domination era and a late time de Sitter expansion. The negative cosmological constant sector of the quantum model is studied in \cite{QC-Letter}. 

\subsection{Singularity avoidance}
A central question in any quantum treatment of cosmological scenario is the status of the big-bang singularity \cite{Thebault:2022dmv}. According to generalized DeWitt's criteria \cite{dewitt_quantum_1967,kiefer_singularity_2019-1}, the quantum model is singularity-free if the probability amplitude for the quantum state vanishes in the vicinity of the classical singularity. In the cosmological scenario, this implies $\mu(v)|\Psi(v)|^2\xrightarrow{v\rightarrow0}0$, where $\mu(v)$ is the measure used to define the inner product. In our case, since $\mu(v)=1,$ $\Psi(v)\xrightarrow{v\rightarrow0}0$ suffice for the singularity avoidance. The near-singularity behavior of the stationary states with a positive cosmological constant is 
\begin{align}
    \lim_{v\rightarrow0}\psi_{k}^1(v)=\lim_{v\rightarrow0}\left(\sqrt{\frac{4\rho_0e^{\frac{2\pi\rho_0}{3k}} }{3k\sinh\left(\frac{2\pi\rho_0}{3k}\right)}}v+O(v^2)\right)\rightarrow0.
\end{align}
Thus, the stationary states follow DeWitt's criteria \cite{dewitt_quantum_1967}, and any wave packets constructed out of these states are expected to avoid singularity. 
It should be noted that the DeWitt's criteria is, in fact, tied to the self-adjointness of the Hamiltonian operator, and the vanishing of the relevant boundary term ensures that the states vanish at the singularity. Thus, the unitarity in the quantum model leads to a singularity-free evolution.

\section{Evolution of the wave packets and quantum bounce}\label{Sec4}
We construct \emph{normalized wave packets} that describe the quantum dynamics of the universe by superposing stationary states of different cosmological constants. These are weighted by a normalized distribution $A(\Lambda)$, or equivalently, by a distribution $A(k)$ where $k=2\sqrt{\Lambda/3}$ is the wave number:
\begin{align}
  \Psi (v, \tl) &= \int_0^\infty \d k \, A(k) \psi^1_k (v) \exp \left( i \Lambda(k) \tl \right)\nonumber \\
                      &= \int_0^\infty dk \, A(k) \psi^1_k (v) \exp \left( i \frac{3}{4} k^2 \tl \right).\label{WP}
\end{align}
The Dirac delta orthonormality of the stationary states ensures the wave packet defined above to be normalized:
\begin{align}
  \int_0^{\infty} \d v \abs{\Psi(v, \tl)}^2 = 1,
\end{align}
provided the weight function is normalized as
\begin{align}
  \int_{0}^{\infty} \d k A^2(k) = 1.
\end{align}
Verifying whether the norm of the wave packet is conserved over its evolution is a consistency test for the formalism discussed above, as the functional analysis of the orthogonality of Kummer's functions for positive eigenvalues is scarce in the literature. This has been verified numerically. 
\begin{figure*}
  \centering
  \setlength{\tabcolsep}{0pt} 
        \begin{tabular}{ccc}
    \begin{minipage}{0.33\linewidth}
      \includegraphics[width=\linewidth]{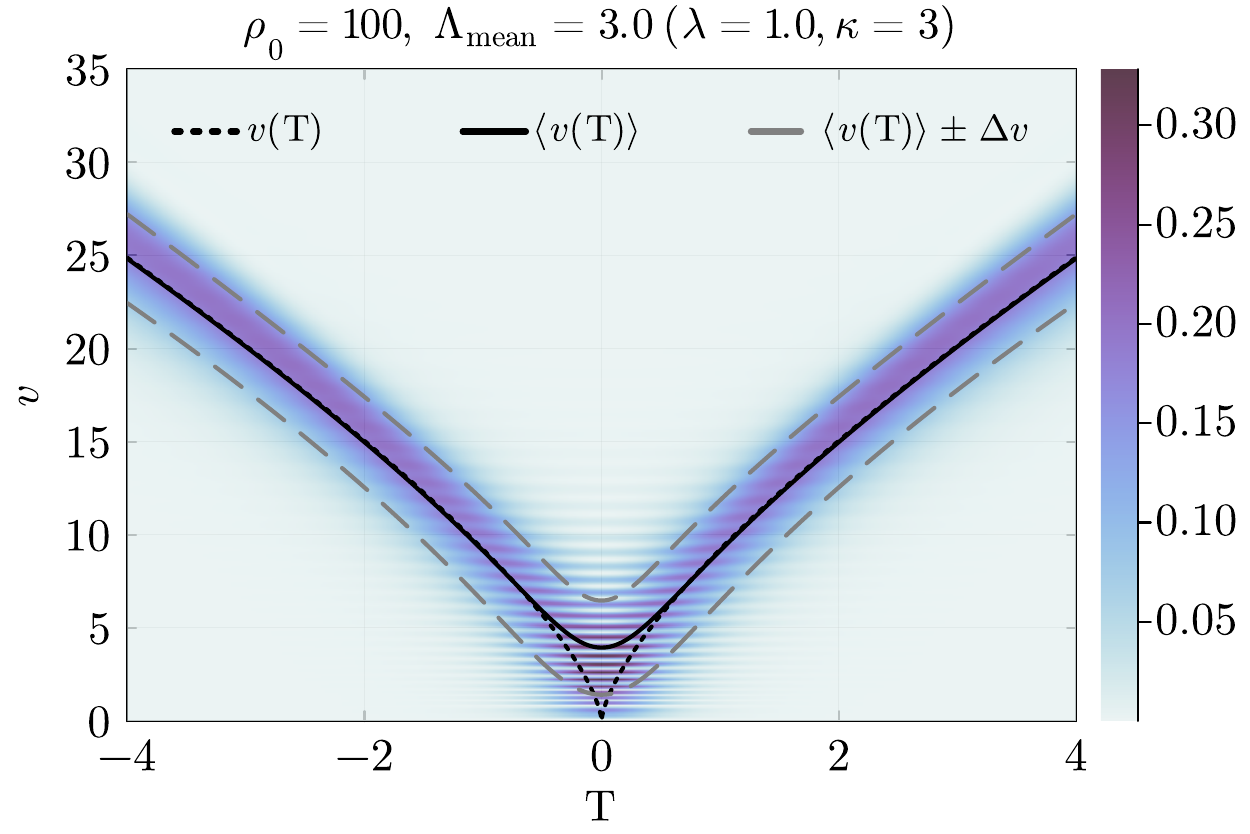}
    \end{minipage} &
    \begin{minipage}{0.33\linewidth}
      \includegraphics[width=\linewidth]{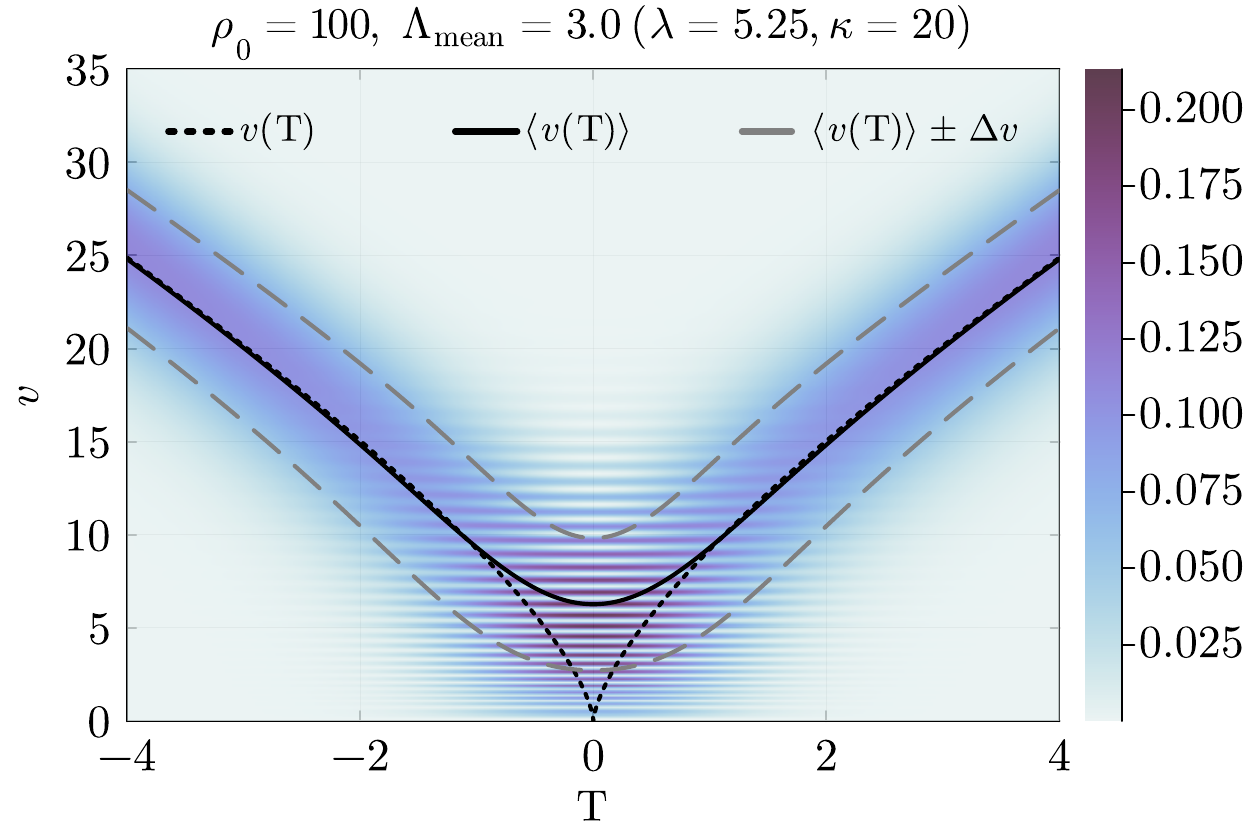}
    \end{minipage} &
    \begin{minipage}{0.33\linewidth}
      \includegraphics[width=\linewidth]{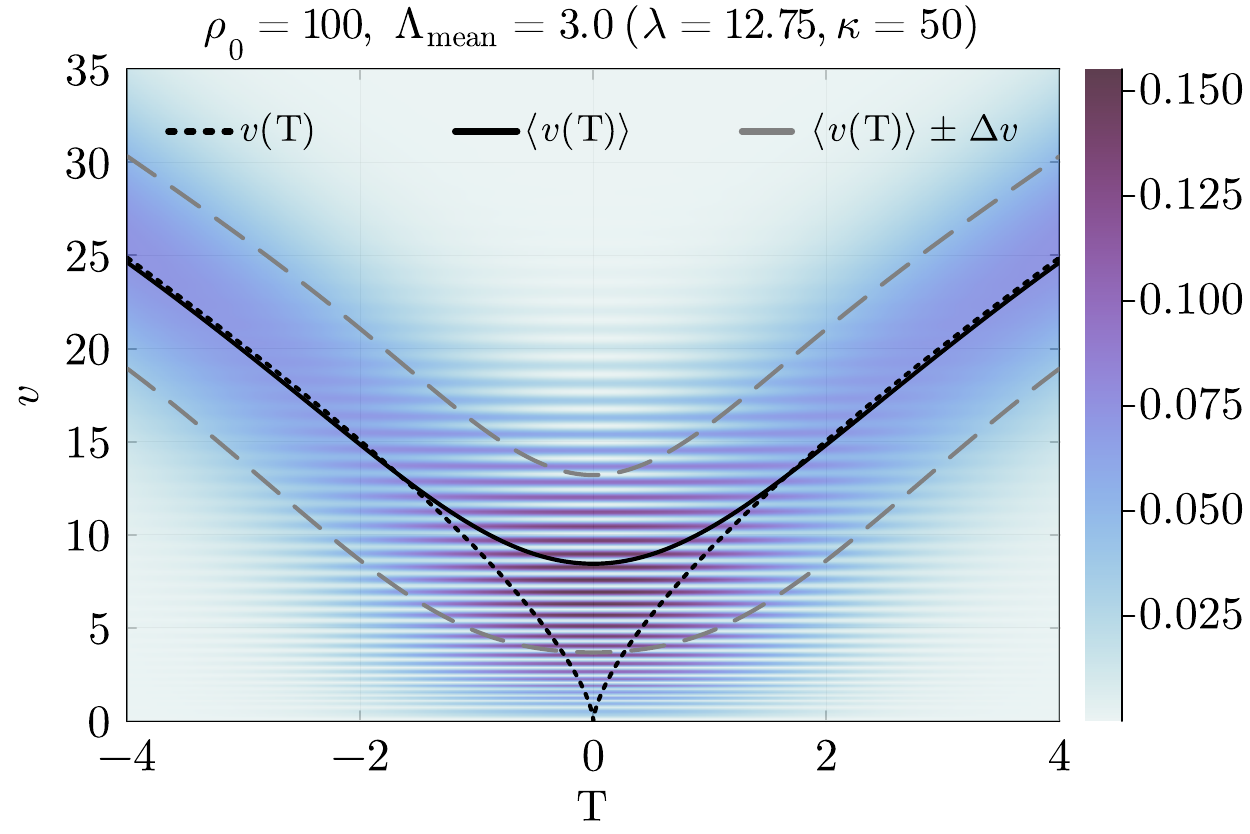}
    \end{minipage} \\
    \begin{minipage}{0.33\linewidth}
      \includegraphics[width=\linewidth]{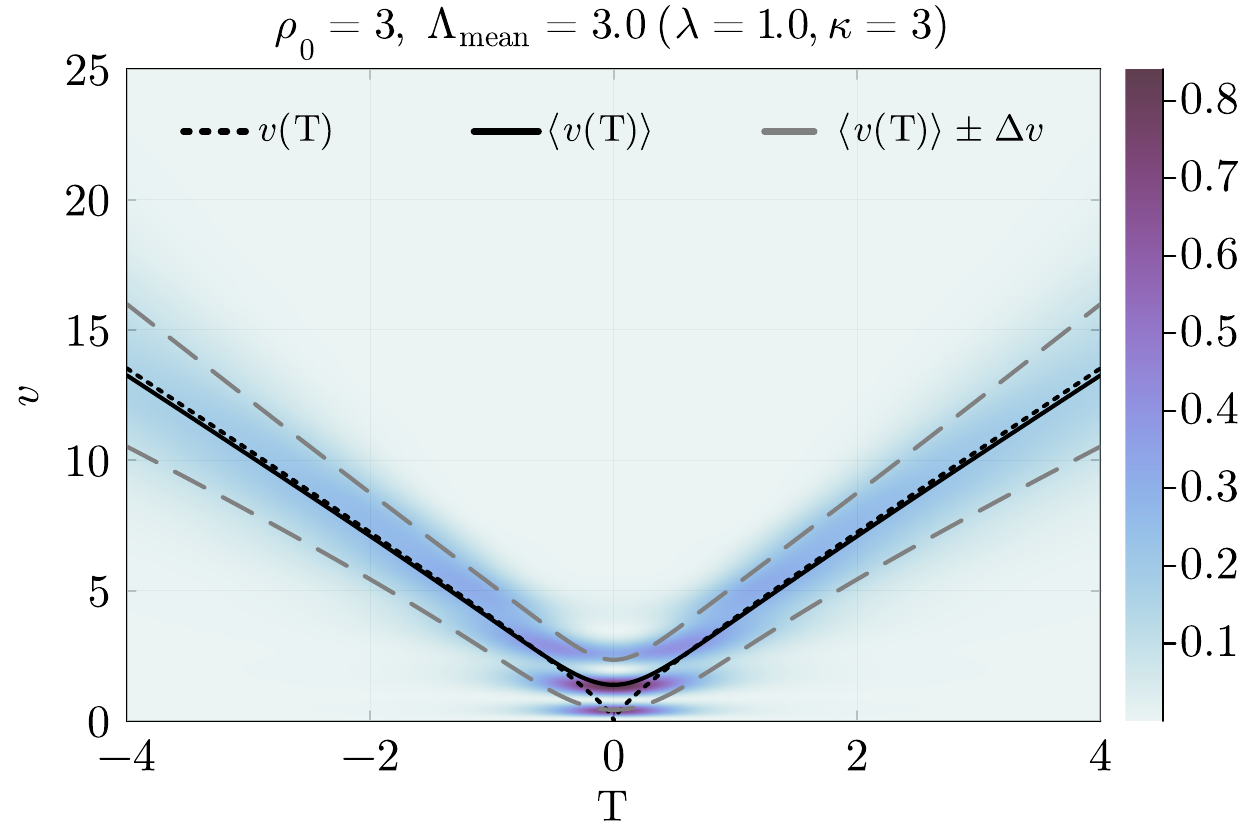}
    \end{minipage} &
    \begin{minipage}{0.33\linewidth}
      \includegraphics[width=\linewidth]{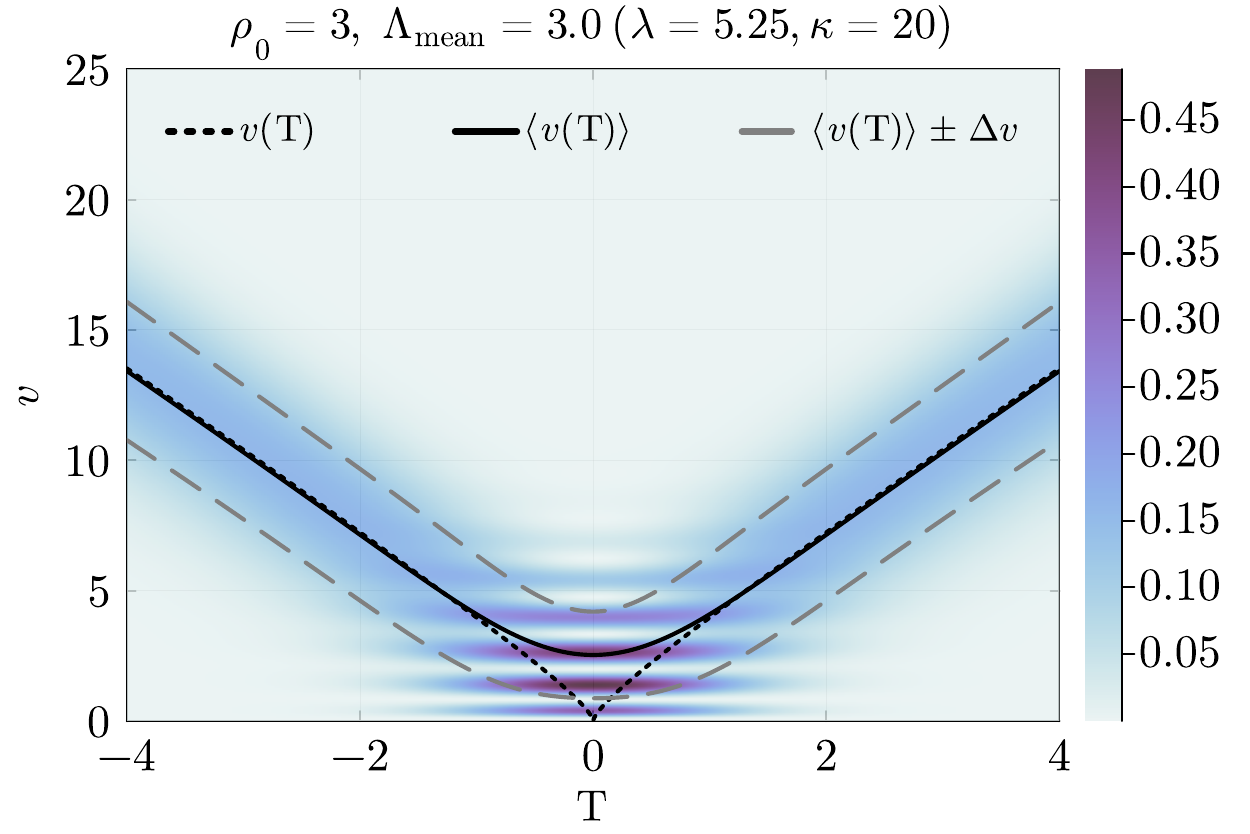}
    \end{minipage} &
    \begin{minipage}{0.33\linewidth}
      \includegraphics[width=\linewidth]{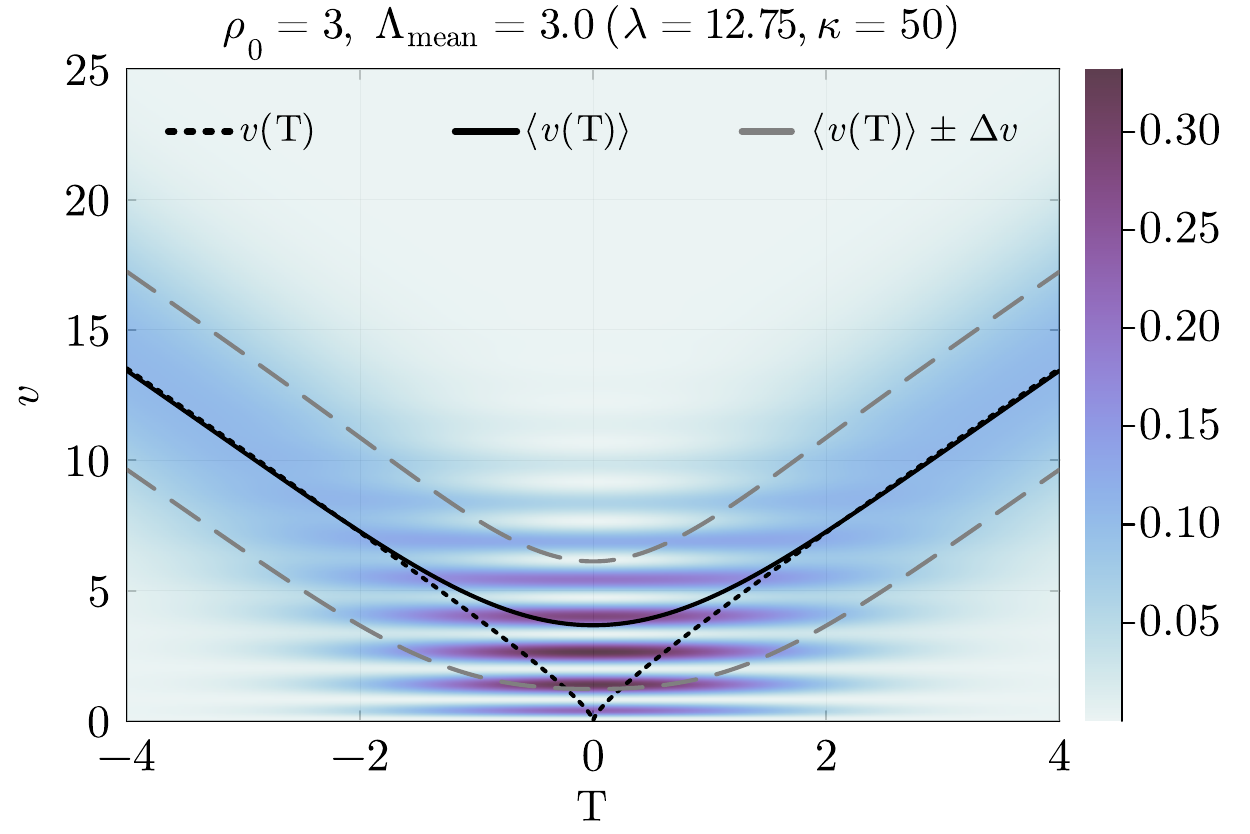}
    \end{minipage} \\
    \begin{minipage}{0.33\linewidth}
      \includegraphics[width=\linewidth]{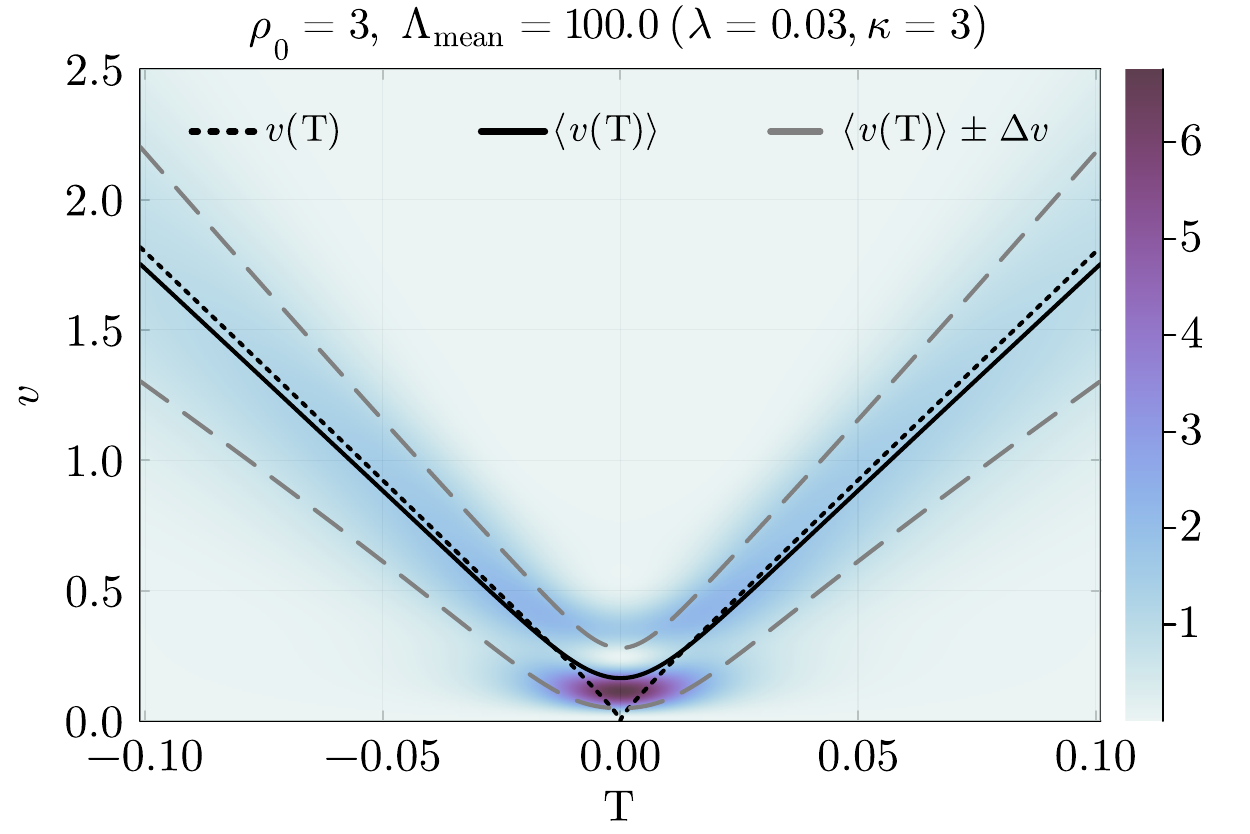}
    \end{minipage} &
    \begin{minipage}{0.33\linewidth}
      \includegraphics[width=\linewidth]{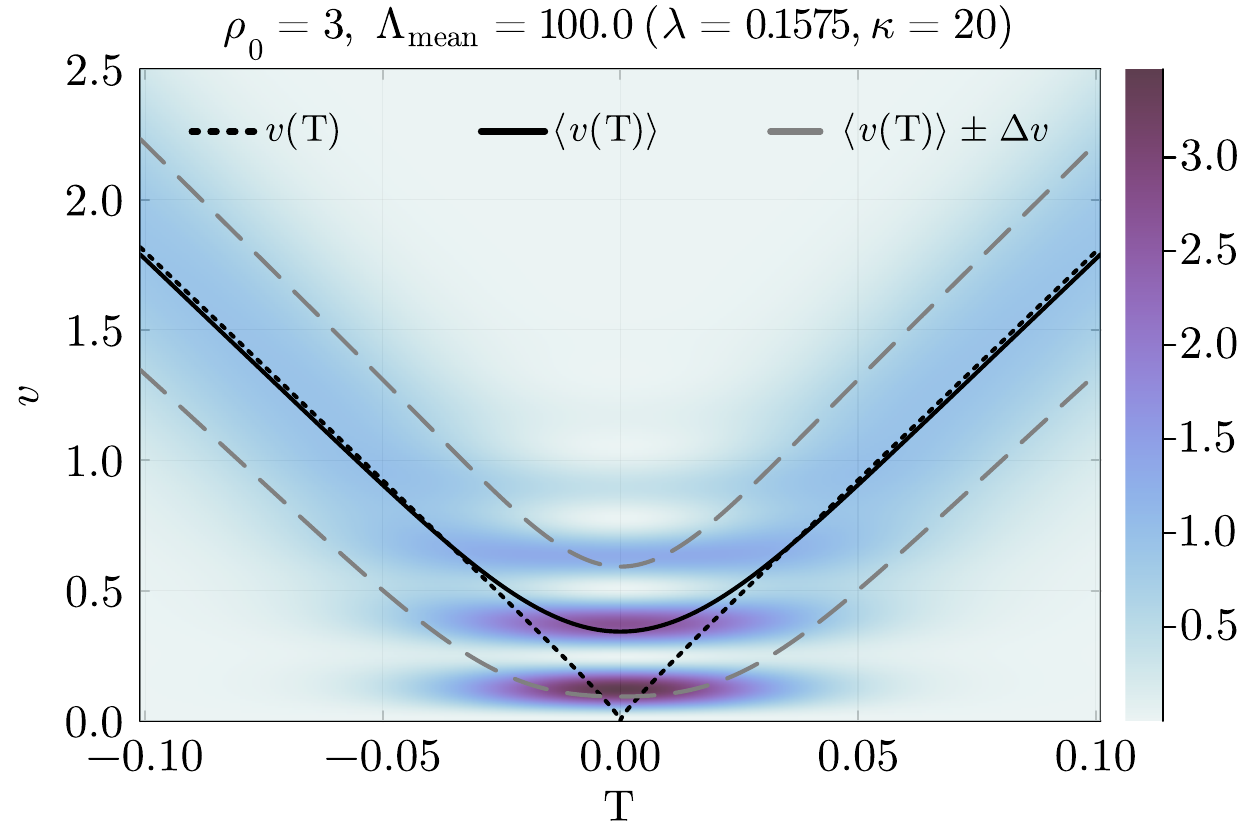}
    \end{minipage} &
    \begin{minipage}{0.33\linewidth}
      \includegraphics[width=\linewidth]{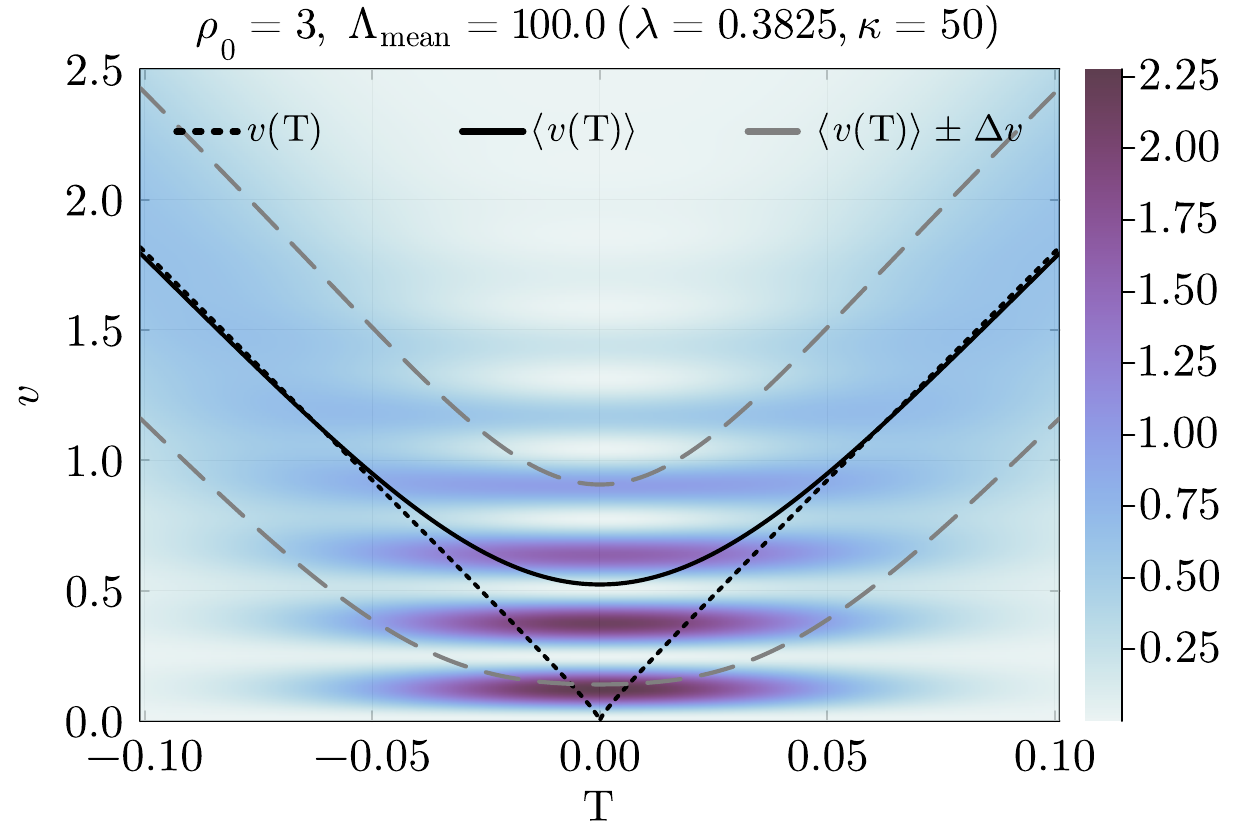}
    \end{minipage} \\
  \end{tabular}
  \caption{Heatmaps for the probability density functions on $(v,\;\tl)$ plane, along with classical trajectories (dotted black line) and the expectation value of the volume variable (solid black line). The gray dashed lines enclose the region of uncertainty $\braket{v}\pm\Delta v$. The first row corresponds to $\rho_0  \gg \Lambda_{\text{mean}}$, the second to $\rho_0 = \Lambda_{\text{mean}}$, and the third to $\rho_0 \ll \Lambda_{\text{mean}}$; while $\kappa$ increases from left to right. The quantities are in reduced Planck units, see Appendix~\ref{Appendix_units}.}
  \label{fig:pdf_grid}
\end{figure*}
In what follows, we consider a Poisson-like distribution as the
weight function, 
\begin{align}
  A(k) = \frac{\sqrt{2} \lambda ^{\frac{\kappa +1}{2}} k^{\kappa
  +\frac{1}{2}} \exp\left(-\frac{k^2 \lambda }{2}\right)}{\sqrt{\Gamma
  (\kappa +1)}},\label{dist}
\end{align}
here $\lambda > 0$ controls the spread and $\kappa > 0$ shapes the skewness. We term this distribution as $\Lambda$-distribution. This choice is motivated by its mathematical tractability and its resemblance to gamma distributions, which naturally describe positive-definite variables like $\Lambda$.
The mean and variance are:
\begin{align}
  \bar{\Lambda} &\equiv \int_{0}^{\infty} \d k \Lambda(k) A^2(k) = \frac{3 ( 1 + \kappa)}{4 \lambda},\\
  \Delta^2_\Lambda &\equiv \overline{\Lambda^2} - \bar{\Lambda}^2 = \frac{9 (\kappa +1)}{16 \lambda ^2} \, ,
\end{align}
yielding a specific variance:
\begin{align}
  \frac{\Delta^2_\Lambda}{\bar{\Lambda}^2} = \frac{1}{1 + \kappa},
\end{align}
independent of the parameter $\lambda$. Thus, larger $\kappa$  leads to a more sharply peaked distribution around $\bar{\Lambda}$, mimicking a classical limit where $\Lambda$ approaches a definite value. In other words, the wave packet represents a quantum superposition of classical universes, each with a distinct $\Lambda$ drawn from \ref{dist}. Consequently, classical observables, like scale factor, Hubble parameter and Ricci scalar, must be interpreted as ensemble averages.  

We numerically evaluate the integral \ref{WP} and obtain the probability density  $|\Psi(v,\tl)|^2$ associated with the wave packet as a function of the volume and clock variables, $v$, $\tl$, for three regimes: $\rho_0 \gg \Lambda_{\text{mean}}$, 
$\rho_0 = \Lambda_{\text{mean}} $ and $\rho_0 \ll \Lambda_{\text{mean}} $, where $\Lambda_{\rm mean} = \bar{\Lambda}$.  \ref{fig:pdf_grid} contains the heatmap for the probability distribution $|\Psi(v,\tl)|^2$ where each row contains the plot for each regime for different $\kappa$ values. 
The classical trajectory is represented by the dotted curves in~\ref{fig:pdf_grid}.
From \ref{fig:pdf_grid}, we infer the following: First, 
the probability distribution has a single-peak profile very early in the collapsing phase ($\tl \to -\infty$). In other words, the probability distribution is peaked on the classical trajectory away from the singularity, enabling us to have a consistent semiclassical picture in this regime \cite{Ashtekar:2005dm}.
Second, as the universe evolves towards the classical singularity, the probability distribution starts to acquire oscillatory features with maximum peaks appearing near the classical singularity ($\tl = 0$). The extent of the oscillatory feature, which appears as the horizontal fringes in the heatmap, depends on the width of the $\Lambda-$distribution. 
For a broadly peaked $\Lambda-$distribution (the first column of~\ref{fig:pdf_grid}), the oscillatory fringes are confined to a small window near the classical singularity. As we decrease the width of the $\Lambda-$distribution, the size of this window increases along with an increase in the frequency of the fringes. 
Third, these oscillatory features start to dissipate as the universe begins to expand after the singularity. The probability distribution is again a single peaked profile as the universe is far away from the bounce. The probability distribution associated with the wave packet vanishes at $v=0$ for all times, showing singularity avoidance according to DeWitt's criteria \cite{dewitt_quantum_1967}. The evolution of wave packet represents the universe tunneling from a collapsing phase to an expanding one, avoiding the classically forbidden singularity.

The oscillatory features near the bounce are typically attributed to the interference of incident and reflected parts of the wave packet \cite{Andrews_WP,ROBINETT20041}, which is sometimes referred to as ``ringing'' \cite{Alexandre:2022ijm,Gielen:2022dhg}. We can divide the evolution of quantum corrected universe in two regimes: semiclassical phases away from the bounce, where the probability distribution is peaked on the classical trajectory, and the quantum phase, where the probability distribution exhibits ringing near the quantum bounce.

Given the numerically constructed wave packets, we compute the expectation value of an operator $\hat{O}$ by numerically estimating the integral:
\begin{align}
    \braket{\hat O}=\int_0^\infty\d v\;\Psi^*(v,\tl)\hat{O}\Psi(v,\tl).
\end{align}
For $\hat{O}=\mathbb{I}$, this integration confirms that the wave packet is normalized, and thus $|\Psi(v,\tl)|^2$ serves as an appropriate probability distribution. We numerically evaluate the expectation values, $\Braket{v}$, and variances, $(\Delta v)^2 = \braket{v^2} - \Braket{v}^2$, of the volume operator for different parameter choices. The numerical estimates of $\Braket{v}$ are plotted as solid black lines in the heatmaps shown in \ref{fig:pdf_grid}, along with their quantum uncertainties $\Braket{v} \pm \Delta v$ (represented by dashed gray lines). The quantum expectation values of the volume variable describe universes undergoing a quantum bounce. 

Classically, a physical singularity at $v=0$ corresponds to a diverging curvature (e.g., $\mathcal{R} \propto 1/a^8 \propto 1/v^{8/3}$ for the given lapse choice) and leads to finite proper-time geodesic incompleteness. By ensuring that the physical volume \emph{$v$ never reaches zero in the quantum description}, our model inherently prevents the associated curvature divergences. Consequently, \emph{proper time geodesics remain complete}, as the quantum corrected spacetime represented by the expectation value of the volume variable in our quantum model never becomes singular. 

\begin{figure}
  \centering
  \includegraphics[width=\columnwidth]{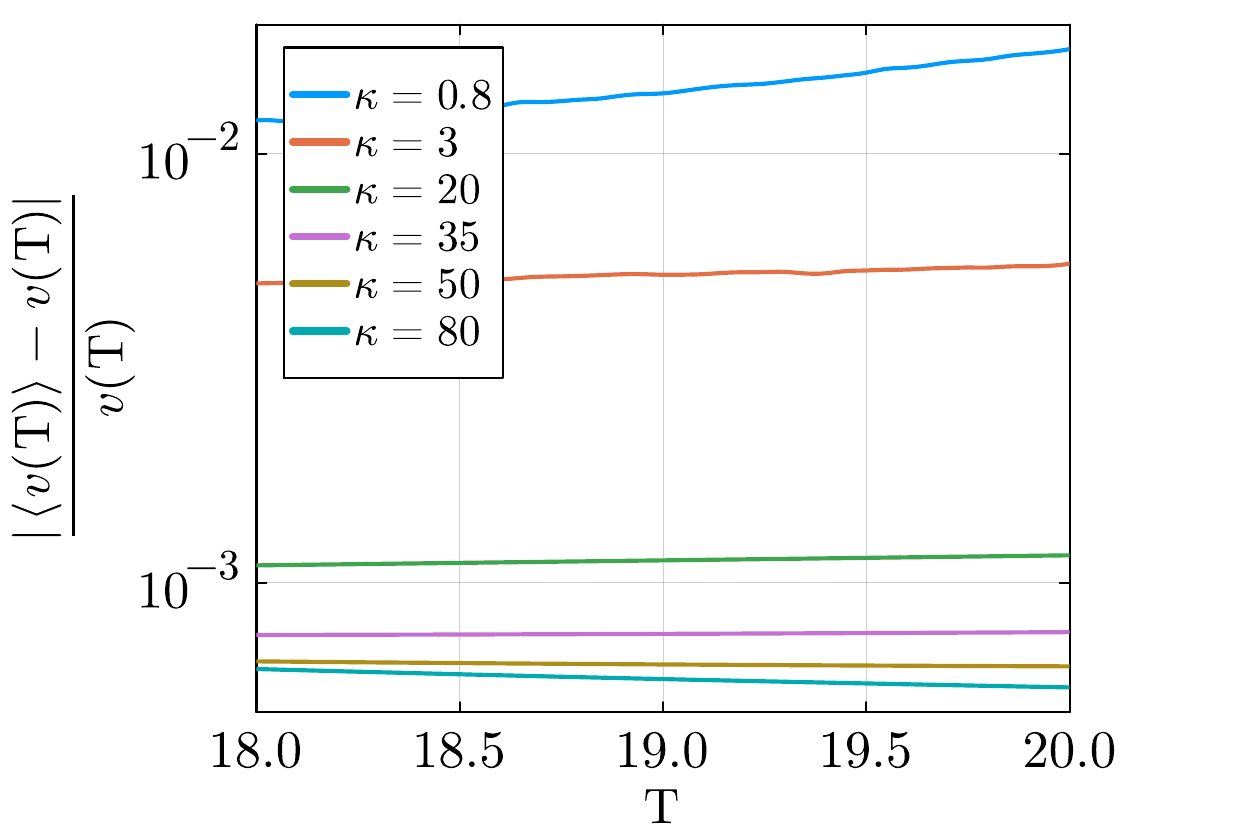}
  \caption{Relative deviation of expectation values of $v$ from the classical trajectory for different $\kappa$ at large values of $\tl$ for $\rho_0=100$ and $\Lambda_{\rm mean}=3 \;(\lambda=1,\;\kappa=3)$. The quantities are in reduced Planck units, see Appendix~\ref{Appendix_units}.}
  \label{fig:largeTl}
\end{figure}

As we transition from the regime $\rho_0\gg\Lambda$ to $\Lambda\gg\rho_0$, $\Braket{v}$ does not asymptote to the classical trajectories for broadly peaked $\Lambda$-distributions (as shown in the first column of \ref{fig:pdf_grid}). Even in the case where $\rho_0\gg\Lambda$, $\Braket{v}$ constructed from border $\Lambda$-distributions (characterized by smaller $\kappa$) deviates from the corresponding classical trajectories, although this is not immediately apparent from \ref{fig:pdf_grid}. For a clearer illustration, the deviation of $\Braket{v}$ from the classical trajectories far from the bounce is shown in \ref{fig:largeTl}. As the width of the $\Lambda$-distribution decreases, the relative deviations of the expectation values from their corresponding classical trajectories diminish. This agreement arises because, for a sufficiently sharply peaked $\Lambda$-distribution, $\overline{f(\Lambda)} \approx f(\overline{\Lambda})$, where the overbar represents the distributional average.

As can be observed from \ref{fig:pdf_grid}, away from the bounce, the region of quantum uncertainty, $\langle v\rangle\pm\Delta v$, captures the spread in the wave packet. However, the oscillations in the probability density near the bounce extend beyond this region $\langle v\rangle\pm\Delta v$. We also observe that for broader $\Lambda$-distributions, both the spread of the probability distribution and the standard deviation in the volume operator increase as we move away from the bounce region (this feature is most visible in the last two plots of the first column in \ref{fig:pdf_grid}).

Furthermore, as we transition from broader to sharper $\Lambda$-distributions, the spreads in the wave packet and quantum uncertainties appear to increase. Note that this feature contradicts the trend observed for the relative fluctuations in the quantum de Sitter universe \cite{Sahota2023,Sahota:2023kox}. Therefore, it is prudent that we thoroughly analyze the quantum uncertainty in the model under consideration. 
\begin{figure*}[t]
  \centering
  \begin{tabular}{c c}
    \includegraphics[width=0.48\textwidth]{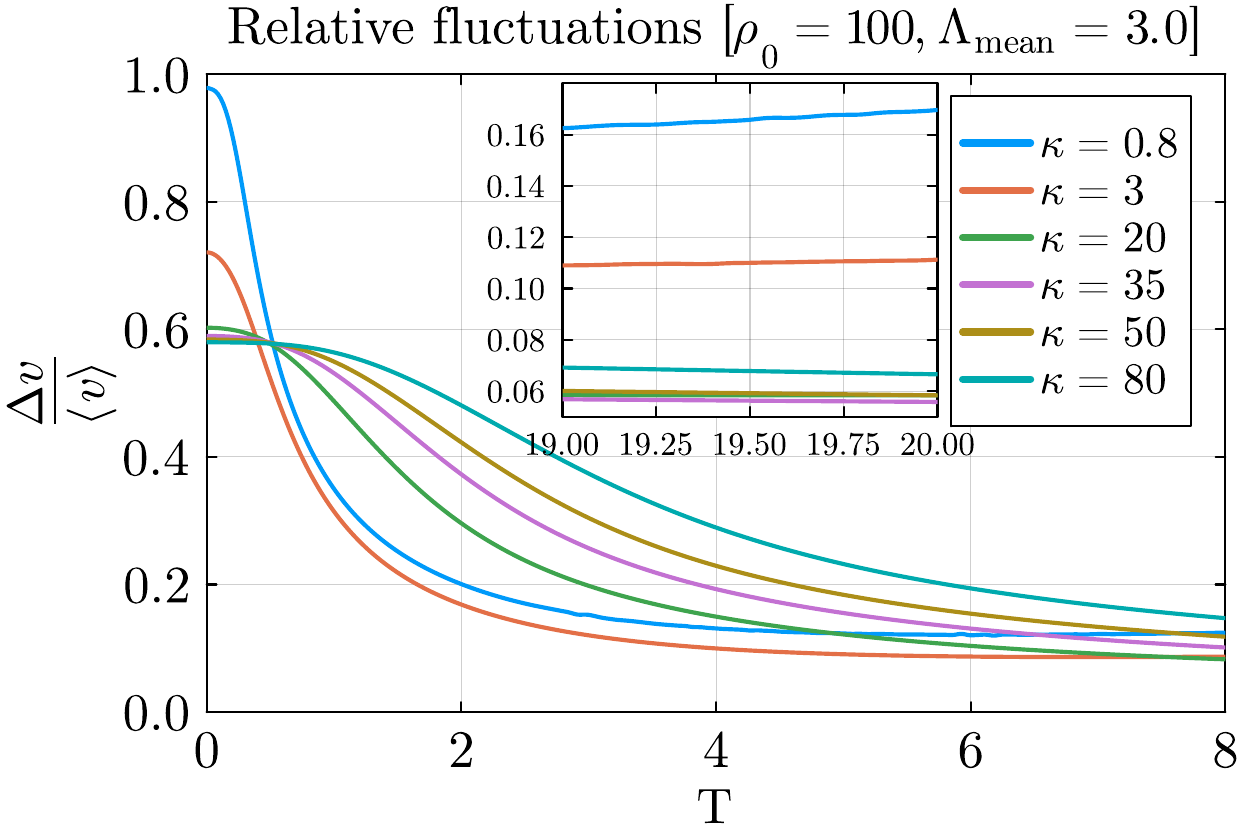} &
    \includegraphics[width=0.48\textwidth]{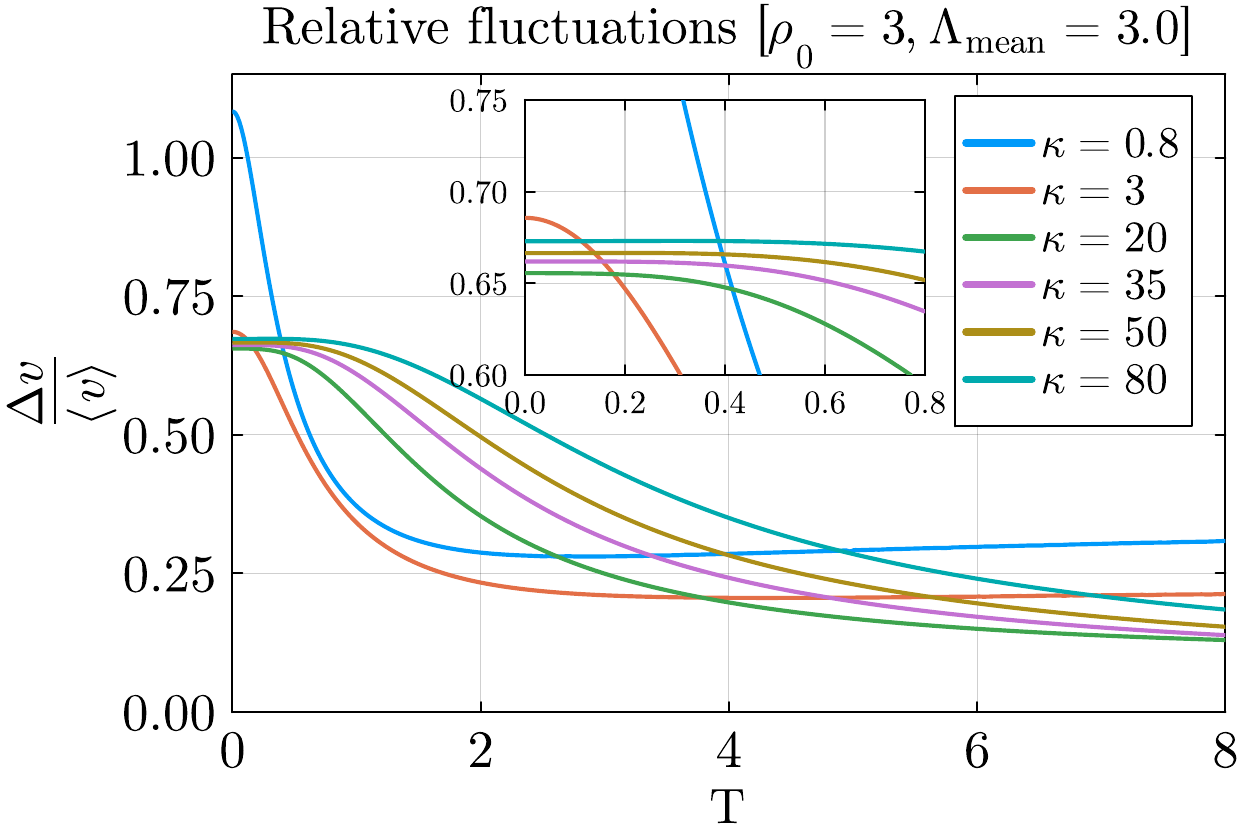}
  \end{tabular}
  \caption{Quantum fluctuations in $v$ for different values of $\kappa$:
    top panel shows: $\rho_0=100.0$; bottom panel: $\rho_0=3.0$ with $\Lambda_{\rm mean}=3$ for both cases. The quantities are in reduced Planck units, see Appendix~\ref{Appendix_units}.}
  \label{fig:delv}
\end{figure*}

To understand the quantum character of the universe in more detail, in \ref{fig:delv}, we plot the relative fluctuation in the volume variable, $\Delta v / \Braket{v}$, as a function of the clock parameter $\tl$ for the two cases $\rho_0\gg \bar{\Lambda}$ and $\rho_0\sim\bar{\Lambda}$. We plot by varying the width of the $\Lambda$-distribution. Two clear trends are apparent in the evolution of the relative fluctuations for both cases. First, the relative quantum fluctuations are maximal at the bounce for all $\Lambda$-distributions, and they decrease as the universe expands. Second, the relative fluctuations do not decay to zero away from the bounce; instead, they saturate to a constant value depending on the width of the $\Lambda$-distribution. These trends are similar to those observed in the case of quantum de Sitter universe in Refs.~\cite{Sahota2023,Sahota:2023kox}: finite quantum fluctuations about the classical behavior highlighting ``classical limit problem" of the quantum model. We further observe that for very broad $\Lambda$-distributions ($\kappa = 0.8, 3.0$ in both figures), the relative fluctuations exhibit larger global maxima at the bounce, which decrease sharply away from the bounce. In contrast, for sharper $\Lambda$-distributions (other values of $\kappa$ in both figures), the relative fluctuations have smaller maxima at the bounce, which slowly decrease towards a constant value as $\tl$ increases.

The \emph{exact behavior} of the fluctuations at the bounce has a different $\kappa$ dependence for the two cases considered. For $\rho_0\gg \bar{\Lambda}$, the magnitudes of the maxima of the relative fluctuations at the bounce tend to be larger as $\kappa\rightarrow0$. These maxima decrease monotonically as $\kappa$ increases. For broader distributions ($\kappa = 20$ to $80$ in the figure), the maxima settle at $\Delta v/\langle v\rangle\sim0.6$. For $\rho_0\sim \bar{\Lambda}$, the maxima of the relative fluctuations at the bounce initially decrease as we increase $\kappa$ (for $\kappa=0.8$ to $3$ in the figure); however, upon further increase in $\kappa$, the maxima start increasing as well, albeit marginally. This can be seen in the inset figure of the second plot in \ref{fig:delv}.

Away from the bounce, the behavior of the relative fluctuations is generally inverted, with sharper $\Lambda$-distributions having larger relative fluctuations and, consequently, a larger spread in the wave packet. This can also be seen in the heatmap plots of \ref{fig:pdf_grid}. However, as the universe evolves further, the relative fluctuations for sharper $\Lambda$-distributions fall below the previously saturated curves for broader $\Lambda$-distributions. In the inset figure of the first plot in \ref{fig:delv}, we see that the relative fluctuations for very broad $\Lambda$-distributions ($\kappa = 0.8$ and $3$ in the figure) indeed settle at larger values (though with small positive slopes, implying they will saturate at even higher values). During the time interval considered in the inset figure, the green curve with $\kappa=20$ is crossing over the brown curve with $\kappa=50$, and it will eventually cross over the teal curve with $\kappa=80$ at a later time. We did not present this behavior explicitly due to large numerical instabilities at large $\tl$. The relatively smaller slopes of the bundle of curves ($\kappa =20,35,50$) and the relatively larger negative slope of the teal curve ($\kappa=80$) clearly indicate that the curves with smaller $\kappa$ will eventually cross the curves with larger $\kappa$ further in the future; therefore, the relative fluctuations settle at smaller values for the sharper $\Lambda$-distributions. A similar trend was observed in \cite{Sahota2023,Sahota:2023kox}.

We now analyze the background dynamics of the quantum universe introduced in the previous section Sec. \ref{Sec3}. Our approach involves calculating the Hubble parameters for the \emph{quantum-corrected spacetime}, which we derive using the expectation value of the volume variable. While a comprehensive understanding of quantum observables ideally requires considering their full expectation values (which depend on both the volume and its conjugate momentum), numerical instabilities arising from integrating highly oscillatory functions and their derivatives prevent a direct investigation of these specific quantum cosmological observables here. Instead, our analysis focuses on the behavior of the \emph{quantum-corrected} spacetime metric:
\begin{align}
    ds^2=-\langle v\rangle^{-2} \d t^2 + \langle v \rangle^{2/3} \delta_{ij}\d x^i \d x^j.
\end{align}
\begin{figure*}
  \centering
  \begin{tabular}{c c}
    \includegraphics[width=0.48\textwidth]{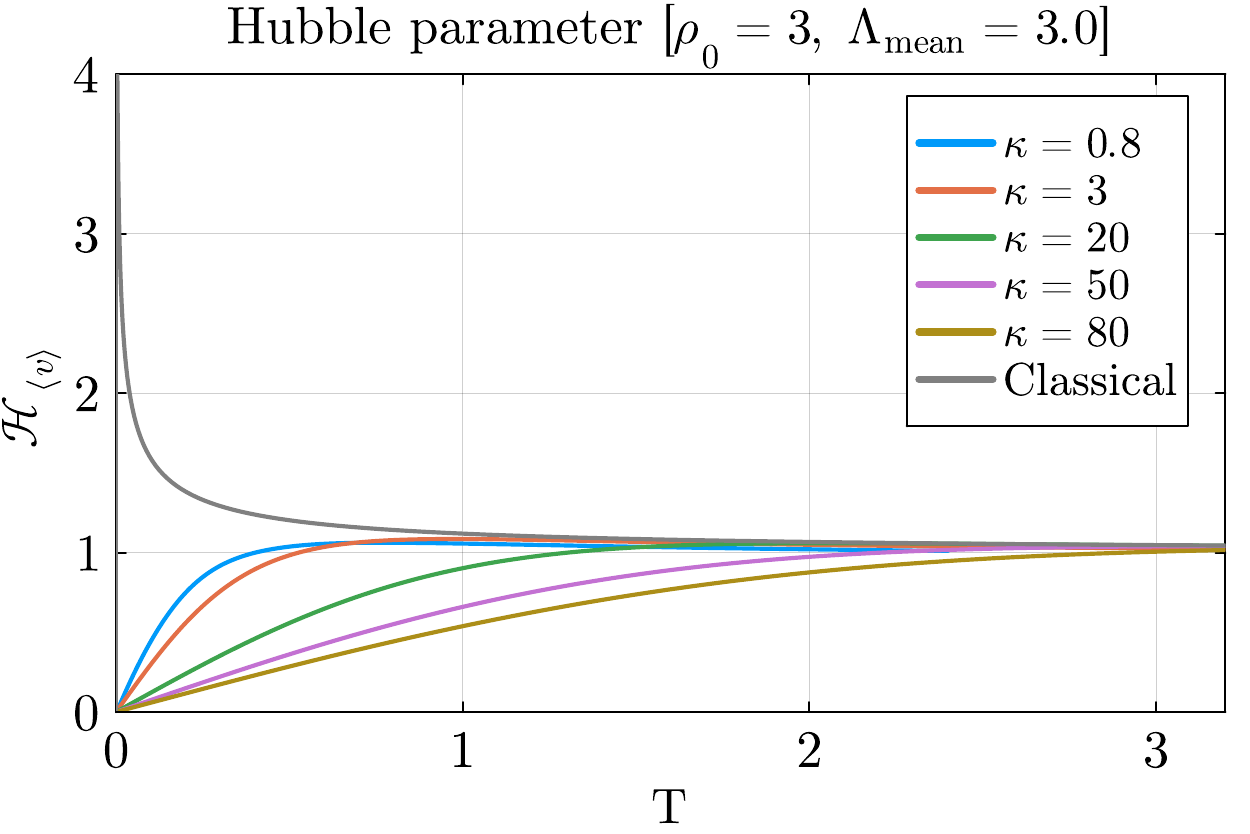} &
    \includegraphics[width=0.48\textwidth]{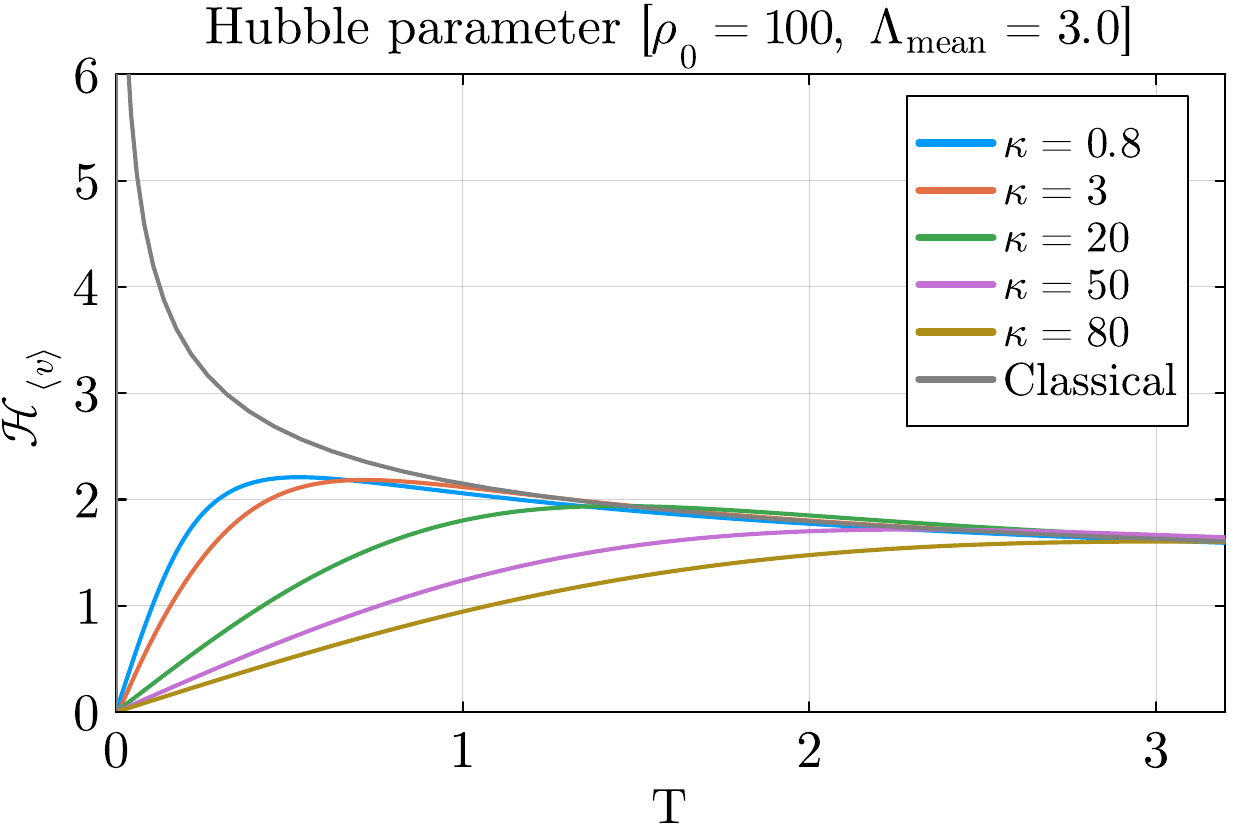}
  \end{tabular}
  \caption{Hubble parameters for different $\kappa$ values:
    top panel shows: $\rho_0=100.0$; bottom panel: $\rho_0=3.0$ with $\Lambda_{\rm mean}=3$ for both cases. The quantities are in reduced Planck units, see Appendix~\ref{Appendix_units}.}
  \label{fig:Hubble}
\end{figure*}
\ref{fig:Hubble} illustrates the Hubble parameter for this quantum-corrected spacetime, comparing it with classical behavior across varying widths of the $\Lambda$-distribution in two distinct scenarios: $\rho_0\gg \bar{\Lambda}$ and $\rho_0\sim\bar{\Lambda}$. Classically, the Hubble parameter diverges at the singularity. In the expanding phase, it monotonically decreases and eventually asymptotes to a finite positive value determined by the cosmological constant ($\sqrt{\Lambda/3}$). 

For the quantum-corrected spacetime, a crucial difference emerges: the Hubble parameter vanishes at the classical singularity, signifying a \emph{quantum bounce}. Away from this bounce, the quantum-corrected Hubble parameters initially increase before eventually converging to the classical value (where $\Lambda = \bar \Lambda$) in regions far from the bounce. Specifically, for larger $\kappa$ values, the Hubble parameters monotonically increase towards the classical value, whereas for smaller $\kappa$, they exhibit a maximum before settling. 

\section{Physical Relevance: The Quantum Matter-Bounce Scenario} \label{Sec5}

A standard assumption in the Hot Big Bang cosmology is that the extremely early universe was radiation-dominated~\cite{Joyce:2014kja}. Consequently, models omitting radiation are sometimes viewed merely as mathematical toy models. However, evaluating the physical significance of a dust-dominated quantum bounce requires situating it within the broader landscape of early-universe paradigms, specifically the matter-bounce scenario~\cite{2002-Finelli.Brandenberger-PRD,Peter:2002cn,Peter:2006hx,Peter:2008qz,Novello:2008ra,Cattoen:2005dx, Battefeld:2014uga, Nojiri:2017ncd,Ijjas:2018qbo,Brandenberger:2016vhg,Malkiewicz:2020fvy,Martin:2021dbz,Martin:2022ptk}.

The matter-bounce paradigm is a well-studied alternative to the cosmological inflation. It resolves the horizon problem and generates a nearly scale-invariant spectrum of primordial curvature perturbations~\cite{1998-Wands-PRD, 2002-Finelli.Brandenberger-PRD}. Crucially, to produce this scale-invariant spectrum, the universe must be dominated by pressureless matter (dust, $w=0$) during the pre-bounce contracting phase. 

In classical approaches to the matter-bounce scenario, avoiding the Big Bang singularity requires introducing exotic physical mechanisms. To transition from contraction to expansion, the system must violate the Strong Energy Condition (SEC) of General Relativity, typically achieved by introducing ghost fields, modified gravity theories (like $f(R)$ or loop quantum cosmology effective dynamics)~\cite{Cheung:2016vze,Barca:2021qdn,Wilson-Ewing:2012lmx}, or exotic fluids~\cite{Ilyas:2020qja}. 

The exact Wheeler-DeWitt solution derived in this work provides a rigorous, first-principles resolution to this problem: First, our model demonstrates that a dust-dominated universe undergoes a non-singular bounce purely due to quantum gravitational effects. The self-adjointness of the Hamiltonian and the resulting wave packet dynamics naturally prevent the volume variable from reaching $v=0$. This provides a fully unitary quantum bounce for the matter-dominated background without requiring any ad-hoc classical SEC violations or exotic modified gravity terms. 

Second, within this paradigm, the fact that our exact solution relies on dust rather than radiation is not a limitation, but a requirement. The $1/v$ effective potential—which maps the system to the hydrogen atom—is generated specifically by the dust energy density. This exact solvability provides a clean, undisturbed background upon which quantum cosmological perturbations can be systematically studied. 

Finally, for a complete cosmological history, a successful matter-bounce scenario requires a period of entropy generation where the universe transitions from matter domination to radiation domination after the bounce~\cite{Brandenberger:2016vhg}. In the context of our model, the quantum bounce occurs in the dust-driven regime. The subsequent inclusion of radiation can be treated analytically as a perturbative correction to the exact hydrogenic wavefunctions once the universe enters the post-bounce expanding phase. Therefore, the exact quantum solutions presented here do not just serve as a mathematical correspondence; they provide a consistent, unitarily evolving background spacetime for the matter-bounce paradigm. By establishing that a dust-dominated universe naturally bounces under standard WDW quantization, this framework offers a concretely solvable foundation for exploring structure formation, primordial perturbations in non-inflationary cosmologies and potential implications for differentiating the bounce from inflation in the future experiments~\cite{Chandran:2024utf}.

\section{Results and discussions}\label{Sec6}

In this work, we have investigated a quantum cosmological model of a spatially flat FLRW universe containing pressureless dust and a cosmological constant, employing the WDW approach to quantum gravity~\cite{wheeler_superspace_nodate,dewitt_quantum_1967,Narlikar:1983lum}.  We modeled the dust fluid using the Brown-Kucha\v{r} formalism \cite{brown_dust_1995} and adopted the unimodular approach to treat the cosmological constant as a dynamical variable, introducing a conjugate time parameter $\tl$ \cite{Unruh:1988in,Henneaux:1989zc}. 
Owing to the inherent lack of a preferred time in quantum gravity, we employed a relational dynamics framework, allowing either the dust fluid's temporal variable ($\td$) or the unimodular variable $\tl$ to serve as a reference clock \cite{Isham:1992ms}. %

A central finding of this work is that the quantum model with unimodular time is exactly solvable due to a striking mathematical analogy with the quantum hydrogen atom. In this correspondence, the volume of the Universe maps to the atomic radial coordinate, the dust energy density parameter ($\rho_0$) to the electrostatic potential strength, and the cosmological constant ($\Lambda$) to the energy of the hydrogen atom. Consequently, states with a positive cosmological constant ($\Lambda > 0$) mirror the scattering states of a hydrogen atom, featuring a continuous spectrum. While both clock choices were considered, the unimodular dynamics framework (using $\tl$) was primarily adopted for detailed analysis, partly due to potential challenges in establishing straightforward eigenstate orthonormality with the dust clock [see Appendix~\ref{Appendix_unimodular}]. The earlier numerical analyses with dust clock choice \cite{Amemiya_2009,maeda_unitary_2015,Ali:2018vmt} show that the quantum model avoids the big bang singularity.

The stationary states satisfy DeWitt's criterion (vanishing at the classical singularity $v=0$), and the quantum model is expected to resolve the big bang singularity. Because the wave function always vanishes at $v=0$, the system effectively undergoes a \emph{quantum turnaround} or bounce in physical volume, preventing geodesic incompleteness. To explore the time evolution of the system and singularity resolution further, we numerically constructed and evolved wave packets using a Poisson-like distribution for the cosmological constant $\Lambda$.

The wave packet dynamics depict a universe undergoing a quantum bounce, replacing the classical singularity. Far from this bounce, the evolution of the wave packet is predominantly semiclassical, with its probability distribution sharply peaked along the classical trajectory. However, near the bounce region, the quantum evolution significantly deviates from its classical counterpart, exhibiting an oscillatory pattern, termed `ringing'. This feature, often attributed to interference between incident and reflected components of the wave packet \cite{Andrews_WP,ROBINETT20041}, shows a dependence on both the mean and the width of the initial $\Lambda$-distribution: the extent and frequency of the ringing become more pronounced for narrower distributions.

Further analysis of expectation values shows that $\braket{v}$ traces a bouncing trajectory, asymptotically matching classical behavior far from the bounce. An effective Hubble parameter derived from $\braket{v}$ corroborates this quantum bounce scenario. While the uncertainty region $\braket{v}\pm\Delta v$ generally captures the spread of the wave packet during semiclassical phases, the ringing features can extend beyond this region. The relative volume fluctuations ($\Delta v / \braket{v}$) are maximal at the bounce, decrease in the approach to the semiclassical regime. Interestingly, the relative quantum fluctuations do not decay to zero but instead settle to a constant non-trivial value far from the bounce, hinting at persistent quantum effects and exhibiting the \emph{classical limit problem}. Similar behavior has also been observed in single fluid quantum cosmology studies~\cite{Sahota2023,Sahota:2023kox}. In contrast, the analysis of quantum fluctuations in the connection-representation theories has appropriate classical limit, as the relative uncertainties in the scale factor vanish~\cite{Alexandre:2022npo}. This disagreement between the metric-representation and connection-representation theories is an intriguing observation, whose origin will be investigated in the future work. The relationship between the width of the $\Lambda$-distribution and relative fluctuations is also subtle, with broader distributions generally leading to larger fluctuations at the bounce and in the semiclassical regime, but with a possible inversion of this trend in intermediate regions.

A natural next step for the present setup would be to include relativistic matter (radiation), however, that would alter the potential in the WDW equation. For a universe with radiation, the effective potential term in the WDW equation typically takes a different form ($V(a) \propto a^{-4}$, depending on the coordinates) compared to the dust-plus-$\Lambda$ case. This change in the potential would \emph{break the direct mathematical analogy to the hydrogen atom}. While the resulting WDW equation might still be analytically solvable in terms of other known special functions, it would not maintain the unique hydrogen-atom-like bound state and scattering solutions that are central to our work. Therefore, extending our exact hydrogen atom analogy to relativistic matter is not straightforward; although the mapping is exact for the current (late) Universe.

Furthermore, the anisotropic singularities are generally more severe than isotropic ones, often exhibiting chaotic Mixmaster behavior as $a \to 0$ (the BKL instability)~\cite{Belinsky:1970ew,Belinski:1973zz}. However, the presence of a \emph{cosmological constant ($\Lambda$) plays a crucial role in suppressing these anisotropies}. As it has been argued previously~\cite{Wald:1983abc,Turner:1986abc}, for a positive $\Lambda$, any initial anisotropies decay exponentially rapidly, ensuring that the universe quickly approaches homogeneity and isotropy at late times. While this doesn't resolve the anisotropic nature of the singularity itself, it provides a strong physical justification for focusing on the isotropic FLRW background when studying the universe's large-scale and late-time quantum dynamics, which is where our correspondence principle applies most strongly. Extending the exact hydrogen atom mapping to anisotropic models, such as Bianchi types, would significantly increase the complexity of the WDW equation (e.g., introducing additional degrees of freedom and cross-terms in the potential), making it highly unlikely to preserve the clean hydrogen atom analogy. It would be interesting to explore whether the presence of cosmological constant does suppress the anisotropies in the quantum cosmological model.

From a phenomenological perspective, the dust-dominated framework studied here has direct physical relevance to the matter-bounce scenario --- an alternative to cosmological inflation. A fundamental requirement of the matter-bounce paradigm is a pressureless, dust-dominated contracting phase, which is necessary to generate a scale-invariant spectrum of primordial perturbations. Because our exact quantum solutions provide a natural, unitarily evolving bounce mechanism without invoking ad-hoc classical violations of the Strong Energy Condition, they establish a rigorously defined background spacetime. Consequently, this exactly solvable model offers a concrete foundation for future work exploring structure formation and the detailed evolution of primordial cosmological perturbations across the quantum bounce in non-inflationary cosmologies.

\section*{Acknowledgments}
HSS and DM thank Kinjalk Lochan for his helpful comments and suggestions. 
HSS is thankful to Suprit Singh for helpful comments. DM is thankful to Shiv K.~Sethi for interesting discussions. The research of HSS is supported by the Core Research Grant CRG/2021/003053 from the Science and Engineering Research Board (SERB), India. DM thanks Raman Research Institute for support through postdoctoral fellowship. HSS's and DM's visits to IIT Bombay was supported by SERB-CRG grant CRG/2022/002348 of SS.

\appendix
\section{WdW equation in natural units}\label{Appendix_units}
In this appendix, we discuss the physical implications of the choice of natural units in the main text, $\hbar=1$, $c=1$ and $8\pi G=1$. Let us consider the total action  
\begin{equation}
S=S_{\mathrm{M}}+S_{\mathrm{G}} \, ,
\end{equation}
where $S_{\mathrm{M}}$ is the matter action, and
\begin{equation}
S_{\mathrm{G}}=\frac{1}{16 \pi G} \int \d^{4} x \sqrt{-g}
\left( \mathcal{R} - 2\Lambda \right)
\end{equation}
Note that we have set $c = 1$ which means space and time have the same units. 
We will also set $\hbar = 1$, hence $G \sim [L]^2$ and the above action is dimensionless. 

To understand $8\pi G=1$ units in the classical system, we introduce a dimensionless time $\tilde{t}=t/\sqrt{8\pi G}$, where $\sqrt{8\pi G}$ is the `reduced' Planck time. In this units, the Friedmann equation for a spatially flat FLRW universe takes the form
\begin{align}
    \frac{\d{a}^{2}}{\d\tilde t^2}&=\frac{1}{3}\left(\tilde\rho a^{2}+\tilde\Lambda a^4\right)
\end{align}
where $\tilde\rho$ and $\tilde\Lambda$ are the dimensionless matter energy density and cosmological constant:
\begin{align}
    \tilde\rho=\rho(8\pi G)^2,\quad\tilde\Lambda=8\pi G \Lambda.
\end{align}
Therefore, in the classical trajectories presented in the main text, energy density, cosmological constant, and the clock parameters are given in their corresponding reduced Planck units. 

Further, re-introducing the $8 \pi G$ factor and the three dimensional spatial volume $V_3$ explicitly, the WDW equation in \ref{eq:WdW}  takes the form
\begin{align}
\left[  - \frac{P_v^2}{4} 
+ \left(\frac{ V_3}{8 \pi G} \right)^2 
\left( \frac{8 \pi G}{3} \rho_0 v^{-1} + \frac{\Lambda }{3}\right)\right]\psi(v)=0.
\end{align}
To understand what $8\pi G=1$ and volume of the spatial section $V_3=1$ means at the quantum level, let us now define the following dimensionless parameters:
\begin{align}
\alpha^{2}  \equiv \frac{ V_3^{2/3}}{8 \pi G } ,\quad
\gamma  \equiv 8 \pi G~ V_3^{2/3} \rho_{0}, \quad
\mu & \equiv V_3^{2/3}~\Lambda,
\end{align}
These dimensionless parameters for a spatial cubic cell of length $l$, i.e., $V_3=l^3$ and with ``reduced" Planck length $l_P=\sqrt{8\pi G}$ takes the form:
\begin{align}
\alpha^{2}  \equiv \frac{l^{2}}{ l_P^2 } \quad
\gamma \equiv  l_P^2l^2 \rho_{0}  \quad
\mu  \equiv l^2\Lambda.
\end{align}
In the dimensionfull analysis, the energy density and cosmological constant differ by $l_P^2$. In terms of these variables the WDW equation becomes
\begin{align}
\left[-\frac{3}{4} \partial_v^2
+\alpha^{4} 
\left(\frac{{\gamma}}{v}  + \mu \right) \right] \psi(v)=0.
\end{align}
Further rescaling the quantities as $\alpha^ 4 \gamma = \tilde{\gamma}$ and  $\alpha^ 4 \mu = \tilde{\mu}$, the WDW equation becomes
\begin{align}
    \left[-\frac{3}{4}\partial_v^2
+
\left(\frac{\tilde{\gamma}}{v} +  \tilde{\mu} \right) \right] \psi(v)=0.
\end{align}
This is the WDW equation in~\ref{eq:WdW}, which is the starting point of the analysis presented in this work. This exercise shows that to reintroduce the units in the quantum analysis, we need to rescale the physical parameters using the transformations discussed above. Thus the cosmological constant and energy density are rescaled by the ``reduced'' Planck length dependent dimensionless factor $\alpha\sim l/l_{\mathrm{P}}$. We should note, the scaling in cosmological constant and energy density should be compensated by rescaling the respective clock variables while writing the separation ansatz $e^{i(\Lambda \mathrm{T}+\rho_0\mathcal{T})}$. Therefore, the numerical values presented in \ref{fig:pdf_grid}, \ref{fig:largeTl}, \ref{fig:delv}, and \ref{fig:Hubble} are of Planck order.

\section{Emergence of unimodular dynamics in the quantum model}\label{Appendix_unimodular}

In the main text, we have written the solutions of the WDW equation for the parameterized system, which have meaningful physical interpretations only with appropriately defined Hilbert spaces. For the quantization of constraint systems, one introduces the notion of kinematical Hilbert space, where all the phase space variables are quantized and promoted to operators on this enlarged Hilbert space \cite{Hohn:2018toe}. The physical Hilbert space is effectively a subspace of the kinematical Hilbert space, obtained by imposing the constraints or by fixing the gauge \cite{BARVINSKY1993237,Barvinsky_2013}. A consistent inner product on this physical space is crucial for probabilistic interpretation and ensuring the self-adjointness of physical observables~\cite{BARVINSKY1993237,Barvinsky_2013}. For the parameterized system, gauge fixing corresponds to choosing the clock out of the phase space variables, leading to the deparameterized system\footnote{When the gauge for a system with time-reparameterization invariance is fixed, the system is termed as a deparameterized system \cite{Gambini:2012ie}.} that represents the physical phase space \cite{Barvinsky:2013aya}. 

For the case at hand, we have several potential choices for a clock variable: $\td$ (dust time), $\tl$ (unimodular time), and even $v$ itself. The volume $v$, representing the spatial size of the universe, can serve as an internal clock, particularly during phases of monotonic expansion or contraction. However, its global suitability can be complex \cite{Bojowald_2011a,Bojowald_2011}. 
In principle, different clock choices may lead to quantum theories with contrasting physical predictions \cite{Gielen:2020abd}. In what follows, we will consider $\tl$ and $\td$ as possible reference clocks and treat the volume variable as the dynamical quantity.

The deparameterized quantum model with $\tl$ as reference clock is:
\begin{eqnarray}
\label{def:Hu}
    \hat{H}_u&\equiv&\biggr[\frac{3}{4}\frac{\partial^2}{\partial v^2}+v^{-1}\rho_0\biggr]\\
    \hat{H}_u\Psi_{}(v,\tl)&=& i\frac{\partial\Psi(v,\tl)}{\partial \tl}\label{HUni}
\end{eqnarray}
where $\hat{H}_u$ is the unimodular Hamiltonian that generates the dynamics with respect to the unimodular time $\tl$. The Hilbert space of this system is chosen to ensure that the unimodular Hamiltonian operator is symmetric, which is $L^2(\mathbb{R}^+,\d v)\otimes L^2(\mathbb{R},\d\td)$, the space of square integrable functions of volume and dust variables on the positive half-plane on the configuration space. Using the identities of hypergeometric functions, we can establish orthonormality of the generalized eigenstates in this case, as we will see in detail in the Appendix~\ref{Appendix_Norm}.

On the other hand, the deparametrized quantum model for the case of the dust variable $\td$ as the reference clock is:
\begin{eqnarray}
\label{def:Hd}
    \hat{H}_d&\equiv&\left[\frac{3v}{4}\frac{\partial^2}{\partial v^2}+v\Lambda\right]\\
    \hat{H}_d\Psi_{}(v,\td)&=& i\frac{\partial\Psi(v,\td)}{\partial \td}\label{Hdust}
\end{eqnarray}
where $\hat H_d$ is the dust Hamiltonian operator that generates the dynamics with respect to the dust comoving time $\td$, and the unimodular parameter is a dynamical variable in this deparameterization choice. The Hilbert space that ensures the hermiticity of the dust Hamiltonian operator is $L^2(\mathbb{R}^+,v^{-1}\d v)\otimes L^2(\mathbb{R},\d\tl)$. 

However, establishing orthonormality relations 
\begin{align}
    \braket{\Psi_{\rho_0}(v)|\Psi_{\rho_0'}(v)}&\sim\delta(\rho_0-\rho_0')\text{ or }\\
    \braket{\Psi_{\rho_0}(v)|\Psi_{\rho_0'}(v)}&\sim\delta_{\rho_0,\rho_0'}
\end{align}
for the eigenstates $\Psi_{\rho_0}(v)$ across the full spectrum of $\Lambda$ and $\rho_0$ 
can be more challenging than for the $\tl$-clock case. This complexity raises questions about the practical construction of a unitary quantum theory with this clock choice, though it does not definitively preclude it.

Moreover, it is not clear whether the issue with the dust clock ($\td$) is curable --- that is, whether the necessary orthonormality relations can be uncovered, or if the problem signifies a true pathology for this setup. This question demands a careful spectral analysis of the dust Hamiltonian operator, which is outside the scope of this work and will be pursued in a future work. The potential complexities inherent in the $\td$ clock choice motivate a thorough investigation of the quantum theory formulated with the unimodular clock $\tl$. Such a formulation is appealing because the unimodular notion of dynamics then appears naturally in this model of quantum cosmology, with the dust variable concurrently functioning as a dynamical component of the system.

However, the question remains: how to treat the dust variable in this quantum model with unimodular dynamics? If we indeed treat it as the eigenvalue of the momentum operator conjugate to dust variable, we have to be cognizant of the spectrum of this operator. Just like when the physical considerations motivate us to consider half-line domain for the scale factor (or equivalently, the volume variable), the momentum conjugate to the scale factor becomes non-self-adjoint \cite{Bonneau_2001,Gitman2012-ya}, and therefore, it is an ill-defined quantum observable. 
In the same vein, physical considerations demand that the dust energy density $\rho_0$  (the eigenvalue of the operator $\hat{P}_\td = -i\partial/\partial\td$) be non-negative. If $\td$ is defined on $\mathbb{R}$, $\hat{P}_\td$ is self-adjoint with a spectrum $(-\infty, \infty)$. Imposing $\rho_0 \geq 0$ means selecting physical states from a portion of this spectrum. This selection, or any alternative definition of $\td$ (e.g., on a half-line to naturally yield $\rho_0 \geq 0$ via self-adjoint extension choices for $\hat{P}_\td$), requires careful handling to ensure overall consistency of the quantum model.

We, however, bypass these issues by working in an effective description that treats the dust parameter $\rho_0$ as a c-number. This approximation is justified if the system is in a state sharply peaked in $\rho_0$. Consider a general state:
\begin{align}
    \ket{\Psi(v,\td,\tl)}=&\int\limits_{\operatorname{Sp_\Lambda}}\!\! \d\Lambda \!\! \int\limits_{\operatorname{Sp}_{\rho_0}}\!\!\d\rho_0\; A(\Lambda)B(\rho_0)\nonumber \\
    &~~~~~~~~~~~~~~e^{i(\Lambda \tl+\rho_0\td)}\ket{\psi_{\Lambda,\rho_0}(v)},
\end{align}
where $\operatorname{Sp_\Lambda}$ and $\operatorname{Sp_{\rho_0}}$ represents the appropriate spectra of $\hat P_\tl$ and $\hat P_\td$ operators. The full density operator is $\hat{\mathbf{\rho}} = \ket{\Psi}\bra{\Psi}$. The reduced density operator for the $(v,\tl)$ subsystem is obtained by tracing over $\td$: $\hat{\mathbf{\rho}}_{\text{red}} = \operatorname{Tr}_\td[\hat{\mathbf{\rho}}]$. Its matrix elements in the $v$-representation,
\begin{align}
    \langle v | \hat{\mathbf{\rho}}_{\text{red}}(\tl, \tl') | v' \rangle = \int \d\td \Psi(v,\td,\tl) \Psi^*(v',\td,\tl'),
\end{align}
evaluate to:
\begin{eqnarray}
& & \langle v | \hat{\mathbf{\rho}}_{\text{red}}(\tl, \tl') | v' \rangle =
\int \d\rho_0~ (2\pi |B(\rho_0)|^2) \\
& &\int \d\Lambda A(\Lambda)e^{i\Lambda\tl}\psi_{\Lambda,\rho_0}(v) 
\int \d\Lambda' A^*(\Lambda')e^{-i\Lambda'\tl'}\psi^*_{\Lambda',\rho_0}(v') \nonumber
\end{eqnarray}
Let $\ket{\Phi_{\rho_0}(\tl,v)} = \int \d\Lambda A(\Lambda)e^{i\Lambda\tl}\psi_{\Lambda,\rho_0}(v)$. Then the reduced density operator is:
$$\hat{\mathbf{\rho}}_{\text{red}} = \int \d\rho_0 (2\pi |B(\rho_0)|^2) \ket{\Phi_{\rho_0}(\tl,v)}\bra{\Phi_{\rho_0}(\tl',v')}$$
This describes an incoherent mixture of states, each corresponding to a specific $\rho_0$, weighted by $2\pi|B(\rho_0)|^2$. If the distribution $B(\rho_0)$ is sharply peaked around a particular value $\bar{\rho}_0$ (i.e., $2\pi|B(\rho_0)|^2 \approx \delta(\rho_0-\bar{\rho}_0)$), then the system is effectively described by the pure state corresponding to $\rho_0 = \bar{\rho}_0$:
$$\hat{\mathbf{\rho}}_{\text{red}} \approx \ket{\Phi_{\bar{\rho}_0}(\tl,v)}\bra{\Phi_{\bar{\rho}_0}(\tl',v')}$$
Therefore, treating $\rho_0$ as a $c-$number (parameter) is justified when considering states with a sharply defined dust energy density. This approximation simplifies the quantum model to a unimodular quantum system with Hilbert space $L^2(\mathbb{R}^+,\d v)$ for a fixed $\rho_0$, alleviates formal challenges, and facilitates numerical computations. Furthermore, states with sharply peaked distributions often exhibit desirable semiclassical properties \cite{Sahota:2023kox}.

The construction of wave packets, crucial for describing time-dependent quantum dynamics, hinges on whether these eigenfunctions (generalized eigenfunctions for continuous spectra, and standard eigenfunctions for discrete spectra) form a complete orthonormal eigen-basis. The existence of such an eigen-basis is guaranteed if the unimodular Hamiltonian $\hat{H}_u$ (defined by its action $\left[\frac{3}{4}\frac{\d^2}{\d v^2} + v^{-1}\rho_0\right]$ and appropriate boundary conditions on $L^2(\mathbb{R}^+,\d v)$) is self-adjoint. The following Appendix addresses the issue of the self-adjointness of $\hat{H}_u$.

\section{Self-adjoint extension of unimodular Hamiltonian}\label{Appendix_self-adjoint}

We are interested in the self-adjointness of the unimodular Hamiltonian, as given in \ref{HUni}, as this property is fundamental for ensuring unitary evolution in the quantum theory and for utilizing the spectral theorem to construct a complete basis of states. Interestingly, the differential operator for $\hat{H}_u$ is similar in form to operators discussed in Section 8.3 of \cite{Gitman2012-ya} (specifically for their case where $g_2=0$, after appropriate scaling and parameter identification).

To define $\hat{H}_u$ as a symmetric operator, we can initially consider its domain $D_0(\hat{H}_u)$ to be the set of smooth functions with compact support within $\mathbb{R}^+$ (i.e., $\psi \in C_0^\infty(\mathbb{R}^+)$), which are dense in the Hilbert space $L^2(\mathbb{R}^+,\d v)$. For $\phi, \psi \in D_0(\hat{H}_u)$, the operator is symmetric because the boundary term from integration by parts,
\begin{align}
    \braket{\phi|\hat{H}_u|\psi}-\braket{\hat{H}_u\phi|\psi}=\left[\phi^*\frac{\partial\psi}{\partial v}-\frac{\partial\phi^*}{\partial v}\psi\right]_0^\infty
\end{align}
vanishes due to the compact support. A symmetric operator $\hat{A}$ is self-adjoint if its domain $D(\hat{A})$ equals the domain of its adjoint $D(\hat{A}^\dagger)$, and $\hat{A}\psi = \hat{A}^\dagger\psi$ for all $\psi \in D(\hat{A})$. Self-adjointness can be established by examining the deficiency indices $(n_+, n_-)$ of the symmetric operator $\hat{H}_u$. These are $n_+=\textrm{dim}(\textrm{ker}\;(\hat{H}_u^\dagger-W\hat{I}))$ and $n_-=\textrm{dim}(\textrm{ker}\;(\hat{H}_u^\dagger-W^*\hat{I}))$ for a non-real complex number $W$ \cite{Gitman2012-ya}. If $n_+=n_-=0$, the operator is essentially self-adjoint. If $n_+=n_-=n \neq 0$, it admits a $U(n)$ family of self-adjoint extensions. If $n_+\neq n_-$, it has no self-adjoint extensions. Based on generalizations of analyses like that in Ref.~\cite{Baldiotti_2011} for similar operators, we proceed with the understanding of the self-adjoint properties of $\hat{H}_u$.

The Hamiltonian $\hat H_u$ under consideration, for $v \in \mathbb{R}^+$, admits a $U(1)$ family of self-adjoint extensions (implying deficiency indices $(1,1)$), parameterized by $\epsilon \in [-\pi/2, \pi/2]$. These extensions are characterized by specific boundary conditions at $v=0$ imposed on general solutions of the form $\psi=a_1\psi^1_{\Lambda,\rho_0}(v)+a_2\psi^2_{\Lambda,\rho_0}(v)$ (where $\psi^1_{\Lambda,\rho_0}(v),~\psi^2_{\Lambda,\rho_0}(v)$ are two linearly independent solutions to $\hat{H}_u\psi = E\psi$, typically corresponding to confluent hypergeometric functions of the first and second kind, $\mathcal{F}$ and $\mathcal{U}$ respectively). The boundary condition is:
$$a_1\cos\epsilon=a_2\sin\epsilon.$$
Following Ref.~\cite{Baldiotti_2011}, the self-adjoint extension corresponding to $\epsilon=\pi/2$ is particularly convenient. This choice implies $a_2=0$, meaning that the solutions involving the hypergeometric function of the second kind ($\mathcal{U}$) do not contribute to the eigenfunctions of this specific self-adjoint Hamiltonian. This selection is crucial for deriving standard orthonormality relations based on the hypergeometric functions of the first kind ($\mathcal{F}$), which are essential for constructing unitarily evolving wave packets. Other choices of $\epsilon \neq \pi/2$ would include contributions from $\mathcal{U}$ functions \cite{Baldiotti_2011,Gitman2012-ya}.

Crucially, the choice of self-adjoint extension determines the spectrum of $\hat{H}_u$. For the physically motivated extensions (assuming $\rho_0 > 0$), the spectrum of $\hat{H}_u$ consists of:
\begin{itemize}
    \item A continuous spectrum for $E_u < 0$. This corresponds to cases where the cosmological constant $\Lambda = -E_u > 0$.
    \item A discrete spectrum of positive eigenvalues $E_{u,n} > 0$. This corresponds to cases where the cosmological constant $\Lambda_{n} = -E_{u,n} < 0$ is quantized as $\Lambda_{n}=-\frac{\rho_0^2}{3n^2}$, leading to $E_{u,n} = \frac{\rho_0^2}{3n^2}$ for $n \in \mathbb{Z}_+$.
\end{itemize}
This means the quantum theory allows only discrete values for a negative cosmological constant, dependent on the dust energy density parameter $\rho_0$. This quantization of $\Lambda$ when it is negative is a notable feature, analogous to the quantization of energy levels in the hydrogen atom, and has been observed in other quantum gravitational scenarios as well \cite{Gielen_PRL}.

\section{Normalization of stationary states}
\label{Appendix_Norm}
In this appendix we arrive at the correct normalization for the stationary states of the unimodular Hamiltonian.

The stationary states of the positive cosmological constant are given by~\ref{SAPosSS}
\begin{align}
    \psi^1_{k}(v)=C_kv\;e^{-ikv}\mathcal{F} \left(1+i\frac{2\rho_0}{3k},2,2ikv\right),
\end{align}
where the normalization factor $C_k$ must lead to $\delta$-function normalization of the states:
\begin{align}
\int_0^\infty \d v  \, \psi^1_k (v) \psi^1_{k'} (v) = \delta(k - k').
\end{align}
Since the orthonormal properties of Hypergeometric functions scarcely studied, we employ the asymptotic normalization procedure used in~\cite{Landau1981-ej}. 
The confluent hypergeometric function can be written as \cite{GRADSHTEYN1980635},
\begin{align}
  \mathcal{F}(\alpha, \beta, z)
  &= \frac{\Gamma(\beta)}{\Gamma(\beta - \alpha)} (-z)^{-\alpha} \, G(\alpha, \alpha - \beta + 1, -z) \notag \\
  &\quad + \frac{\Gamma(\beta)}{\Gamma(\alpha)} e^z z^{\alpha - \beta} G(\beta - \alpha, 1 - \alpha, z).
\end{align}
In our notation $\alpha = \frac{2i\rho_0}{3k} + 1$, $\beta = 2$, and $z = 2 i k v$.
Using these, the wavefunction can be cast in the form
$\psi_k^1 (v) = C_k \left( \psi_1 + \psi_2 \right)$,
where,
\begin{strip}
\begin{align}
  \psi_1
  &= v\, e^{-i k v}  \frac{\Gamma(2)}{\Gamma(1 - \frac{2i\rho_0}{3k})} (-2 i k v)^{-1 - \frac{2i\rho_0}{3k}}
  G\left( 1 + \frac{2i\rho_0}{3k}, \frac{2i\rho_0}{3k}, -2 i k v \right)\\
  \psi_2 
  &= v \, e^{i k v} \frac{\Gamma(2)}{\Gamma(1 + \frac{2i\rho_0}{3k})} (2 i k v)^{-1 + \frac{2i\rho_0}{3k}} G\left(1 - \frac{2i\rho_0}{3k}, -\frac{2i\rho_0}{3k}, 2 i k v \right).
\end{align}
We see that $\psi_1 = \psi_2^*$, and the total wavefunction becomes
\begin{align}
  \psi^1_k (v)
  =  \frac{C_k } {k\,e^{\pi\rho_0/3k}}
    \operatorname{Re} \biggr\{
     e^{\left( -i \left( k v - \frac{\pi}{2} + \frac{2\rho_0}{3k} \ln 2 k v \right) \right)}
     \frac{G \left(1 + \frac{2i\rho_0}{3k}, \frac{2i\rho_0}{3k}, -2 i k v \right)}{\Gamma(1 - \frac{2i\rho_0}{3k})} \biggr\}.
\end{align}
\end{strip}
We now consider the asymptotic expansion
of $ \psi^1_k(v) $ for large $ v $. Noting that the function
$ G(\alpha, \beta, z) $ can be expanded as
\begin{align}
G(\alpha, \beta, v) &= 1 + \frac{\alpha\beta}{1!z} + \frac{\alpha (\alpha+1) \beta (\beta+1)}{2!z^2} + \dots,
\end{align}
the wave function in the $v \to \infty$ limit takes the form
\begin{align}
\psi_k ^1(v) \approx \frac{C_k } {k\,e^{\pi\rho_0/3k}}  \frac{\sin \left( k v + \frac{2\rho_0}{3k}\ln 2 k v + \delta_k \right)}{|\Gamma(1 - \frac{2i\rho_0}{3k})|},
\end{align}
where $\delta_k=\operatorname{arg}\Gamma(1 - \frac{2i\rho_0}{3k})$.
Thus, for large $v$, the inner product can be written as:
\begin{align*}
\int \d v  \, \psi_k^1 (v) \psi_{k'}^1 (v) =\int \d v\; \frac{|C_k|^2 } {k^2\,e^{2\pi\rho_0/3k}}  \frac{\sin ^2\left( k v \right)}{|\Gamma(1 - \frac{2i\rho_0}{3k})|^2}
\end{align*}
where we have used the orthogonality relation~\cite{Landau1981-ej}
\begin{align}
\int \d r  r^2 \frac{\sin(k_1 r)}{r} \frac{\sin(k_2 r)}{r} = \frac{\pi}{2} \delta(k_1 - k_2).
\end{align}
From this, we obtain
\begin{align}
  C_k
  &= \sqrt{\frac{2}{\pi}} 
  e^{\frac{\pi\rho_0}{3k}} k 
  \left|\Gamma \left(1 - \frac{2i\rho_0}{3k} \right)\right|\\
  &= 2\sqrt{\frac{e^{\frac{2 \pi \rho_0}{3k}}\rho_0}{3k\sinh(2\pi\rho_0/3k)}},
\end{align}
noting $|\Gamma(1+ia)|=\sqrt{\pi a/\sinh(\pi a)}$ \cite{GRADSHTEYN1980635}.


\input{bib_resource.bbl}

\end{document}

%% file: bib_resource.bbl
\providecommand{\href}[2]{#2}\begingroup\raggedright\endgroup